\documentclass[twocol]{ametsocV6.2}
\usepackage{booktabs} 
\usepackage{tabularx} 
\usepackage{float}
\usepackage{etoolbox}
\usepackage{placeins}
\usepackage{balance}
\usepackage{threeparttable}
\usepackage{xcolor}
\usepackage{hyperref}       
\hypersetup{
pdftitle={Coupled Seasonal Data Assimilation of Sea Ice, Ocean, and Atmospheric Dynamics over the Last Millennium},
pdfauthor={Zilu Meng; Gregory J. Hakim; Eric J. Steig},
pdfsubject={Author manuscript; published in Journal of Climate, 38, 7229--7247 (2025)},
pdfkeywords={ENSO; Climate; Climate variability; Paleoclimate; Data assimilation; Multidecadal variability},
colorlinks=true,
linkcolor={blue!50!black},
citecolor={blue!50!black},
urlcolor={blue!50!black}
}
\renewcommand\abstract[1]{\def\theabstract{%
  \centerline{\vtop{\bigfootnotesize\hsize=0.90\textwidth\linewidth=\hsize
  \noindent ABSTRACT:\;\;#1\vskip1sp}}}}
\title{Coupled Seasonal Data Assimilation of Sea Ice, Ocean, and Atmospheric\\Dynamics over the Last Millennium} 
\DeclareRobustCommand{\JCLIpublicationnotes}{%
\par\smallskip\noindent Supplementary Information: \url{https://zilum.github.io/paper/LMR_Season-2.pdf} (\href{https://doi.org/10.1175/JCLI-D-25-0048.s1}{publisher supplement}).
\par\smallskip\noindent Author manuscript: Meng, Z., G. J. Hakim, and E. J. Steig, 2025: \textit{J. Climate}, \textbf{38}, 7229--7247, \url{https://doi.org/10.1175/JCLI-D-25-0048.1}. \copyright\ 2025 American Meteorological Society; \href{https://www.ametsoc.org/PUBSCopyrightPolicy}{copyright policy}. This author-typeset copy may differ from the version of record.}
\authors{Zilu Meng\aff{a}\correspondingauthor{Zilu Meng, zilumeng@uw.edu\JCLIpublicationnotes},
Gregory J. Hakim\aff{a},
and Eric J. Steig\aff{b,a} 
}
\received{Manuscript received 23 January 2025, in final form 1 July 2025, accepted 28 August 2025}
\affiliation{\aff{a}{Department of Atmospheric and Climate Science, University of Washington, Seattle, Washington, USA}\\
\aff{b}{Department of Earth and Space Sciences, University of Washington, Seattle, Washington, USA}\\
}
\abstract{``Online" data assimilation (DA) is used to generate a seasonal-resolution reanalysis dataset over the last millennium by combining forecasts from an ocean--atmosphere--sea-ice coupled linear inverse model with climate proxy records. Instrumental verification reveals that this reconstruction achieves the highest correlation skill, while using fewer proxies, in surface temperature reconstructions compared to other paleo-DA products, particularly during boreal winter when proxy data are scarce. Reconstructed ocean and sea-ice variables also have high correlation with instrumental and satellite datasets. Verification against independent proxy records shows that reconstruction skill is robust throughout the last millennium. Analysis of the results reveals that the method effectively captures the seasonal evolution and amplitude of El Ni\~{n}o events, seasonal temperature trends that are consistent with orbital forcing over the last millennium, and polar-amplified cooling in the transition from the Medieval Climate Anomaly to the Little Ice Age.}
\makeatletter
\renewcommand\footnoterule{\kern-3pt\hrule\@width.2\columnwidth\kern2.6pt}
\patchcmd{\@maketitle}{\vglue61pt}{\vglue24pt}{}{\errmessage{AMS title-spacing patch failed}}
\patchcmd{\@maketitle}{\vskip17pt}{\vskip8pt}{}{\errmessage{AMS abstract-spacing patch failed}}
\g@addto@macro\@maketitle{%
  \begingroup
  \setlength{\fboxsep}{6pt}
  \setlength{\fboxrule}{0.4pt}
  \centerline{\fcolorbox{black!10}{black!10}{%
  \begin{minipage}{\dimexpr0.90\textwidth-2\fboxsep-2\fboxrule\relax}
  \bigfootnotesize\noindent SIGNIFICANCE STATEMENT:\; This paper introduces a new seasonal-resolution reanalysis of the last millennium, based on an ``online" data assimilation method using a linear inverse model to assimilate paleoclimate proxies. We find good agreement when verifying the reconstruction against modern instrumental reanalyses and out-of-sample proxies. Results show that seasonal temperature trends are similar to predictions from orbital-insolation trends, and seasonal variability of modern El Ni\~{n}o events is similar to instrumental reanalyses. This framework offers a dynamically consistent, seasonally resolved view of past climate variability that supports broader applications in paleoclimate research.
  \end{minipage}}}
  \endgroup
  \vskip8pt
  \centerline{\vtop{\bigfootnotesize\hsize=0.90\textwidth\linewidth=\hsize
  \vskip1sp}}
  \vskip10pt
  \centerline{\rule{0.25\textwidth}{0.6pt}}
  \vskip12pt
}
\makeatother

\runningheadauthors{Meng et al.}
\begin{document}
\raggedbottom
\brokenpenalty=10000

\maketitle
\lhead{\small\itshape Author manuscript}
\chead{\small\scshape Meng et al.}
\rhead{\small\thepage}
\cfoot{}

\section{Introduction}

Reconstructions of past climate are essential for understanding the dynamics of the long-term climate system. Such reconstructions are particularly important in the context of global warming \citep{pachauri2007ipcc}, as they place contemporary climate variability within a larger sample of past climate. This historical and long-term perspective also enhances our ability to improve projections of future climate change by providing a reference against which model simulations can be compared. Before the instrumental era, when humans began using scientific tools to record weather and climate information like temperature and precipitation, climate information is primarily derived from natural proxies such as tree rings, corals, and ice cores. For example, the width of some tree rings reflects local moisture and temperature stress \citep[e.g.,][]{briffa2004large}. This information can be used to reconstruct past climate conditions. The challenges of using proxies to reconstruct past climate include their uneven spatial distribution, inconsistent temporal coverage, varying temporal resolution, complex relationships  with physical drivers, and complex noise characteristics \citep[e.g.,][]{christiansen2017challenges}, complicating multiproxy interpretations of climate variability.
Recently, data-assimilation (DA) methods \citep[e.g.,][]{bouttier2002data} have been increasingly used to reconstruct past climates \citep[e.g.][]{dirren2005toward,goosse2010reconstructing, widmann2010using,franke2017monthly,perkins2021coupled, tardif2019last, steiger2018reconstruction,valler2024mode}. This approach combines climate model physical constraints with proxy data to reconstruct climate variables. One of the most significant advantages of DA is that it allows for the reconstruction of variables not directly represented by the proxies \citep{hakim2016last}. For example, we can use temperature data from proxies to infer sea-ice conditions and geopotential height, as there are strong and well-understood  correlations among these variables \citep[e.g.][]{hakim2016last,steiger2018reconstruction,brennan2022reconstructing_inst,brennan2022reconstructing,meng2024reconstructing}.

A general and flexible paleoclimate data assimilation (PDA) framework, the Last Millennium Reanalysis (LMR), was proposed by \citet{hakim2016last} for reconstructing climate variables over the Common Era. The success of this framework has been followed by extensive research on PDA \citep[e.g.,][]{steiger2017, dee2020enhanced, sun2022analog, luo2022last, meng2024reconstructing, zhu2023pseudoproxy, okazaki2021revisiting, hu2024reconstructing}. One practical limitation of PDA compared to weather DA is the high cost of forecasts that generate the prior (``first guess"), because of the need for long integrations of climate models \citep[e.g.,][]{taylor2012overview}. Consequently, the initial LMR framework used an ``offline" data assimilation (DA) method, where the prior is sampled from a static source, such as existing climate model simulations. 
This approach is particularly relevant in paleoclimate applications, where the forecast skill of dynamical climate models may be limited, and the computational cost of running those models is high.  

There are, however, patterns of variability, such as the Pacific Decadal Oscillation (PDO) \citep[e.g.,][]{mantua2002pacific} and El Ni\~{n}o South Oscillation (ENSO) \citep[e.g.,][]{mcphaden2006enso,meng2024pacific}, that persist on seasonal to interannual timescales. Developing ``online" PDA methods that exploit this persistence can lead to more accurate reconstructions, since the memory of past proxies is transmitted into the future by the forecast model. Using DA with a skillful coupled atmosphere--ocean model, information from terrestrial proxies such as tree-ring width can be used to inform ocean state estimates, which then carry memory in the model forecast through ocean persistence. An example of this online PDA approach is shown by \citet{perkins2021coupled}, who used a linear inverse model (LIM) to reconstruct climate fields over the last millennium and found improved representations of decadal variability.

A significant challenge with PDA reconstructions is resolving the seasonal cycle. For example, proxies from Northern Hemisphere trees, including tree-ring width (TRW) and latewood density, primarily reflect warm-season temperature \citep{pages2013continental, pages2k2017global}. Previous PDA studies have used this information to reconstruct annual-mean climate variability, leading to biases in the reconstructions and inconsistent results for significant climate periods such as the Medieval Climate Anomaly and the Little Ice Age \citep{pages2k2017global, hakim2016last, tardif2019last, steiger2018reconstruction}. Here, we present results for a seasonal reconstruction of the last millennium using online PDA. We use a LIM to forecast one season to the next: from March-May (MAM) to June-August (JJA), from JJA to September-November (SON), from SON to December-February (DJF), and from DJF to the next year. 

The LIM incorporates sea-ice variables (concentration and thickness), recognizing the long-lead memory of sea ice \citep{blanchard2011persistence} and therefore predictive skill, especially near the Arctic where seasonal variability is large. We assimilate proxies from the PAGES2k V2 \citep{pages2k2017global} database at the season specific to each proxy. Proxies that represent annual-mean conditions are assimilated subsequently once an annual mean forecast is available from the initial LIM forecast from seasonal proxies only.

The organization of the remainder of the paper is as follows. Section \ref{sec:method} details the PDA methods and data used in this study, and Section \ref{sec:verification} presents instrumental and proxy verification to measure the accuracy of the reconstruction. Section \ref{sec:MCALIA} applies the reconstruction to analyze seasonal climate variability and trends over the last millennium. Section \ref{sec:conclusion} provides a concluding discussion.

\section{LMR-Seasonal Framework Data and Methods} \label{sec:method}

The LMR-Seasonal approach utilizes an online ``cycling" DA framework, consisting of three components, with the goal of reconstructing a subset of atmosphere--ocean--sea-ice fields (Table \ref{tab:recon_vars}). The first component involves a simplified climate model, a LIM trained for seasonal forecasting as described in Subsection \ref{sec:LIM}. The second component, proxy system models, which estimate the proxies from the prior, are trained as described in Subsection \ref{sec:PSM}. The third component is an Ensemble Kalman Filter (EnKF), which is used to combine the proxy and prior as described in Subsection \ref{sec:enkf}. 

\begin{table*}[t]
\centering
\caption{Variables reconstructed in LMR Seasonal and their spatial coverage}
\label{tab:recon_vars}
\begin{tabular}{llc}
\toprule
\textbf{Variable} & \textbf{Description}  & \textbf{Spatial Coverage} \\
\midrule
TAS & 2-meter air temperature   & Global (2° × 2°) \\
TOS & Sea surface temperature   & Global (2° × 2°) \\
OHC300 & Ocean heat content (0–300 m)   & Global (2° × 2°) \\
SIT & Sea-ice thickness (Northern Hemisphere)   & NH  (2° × 2°) \\
SIC & Sea-ice concentration (Northern Hemisphere)   & NH (2° × 2°) \\
\bottomrule
\end{tabular}
\end{table*}

\subsection{Linear Inverse Model (LIM)} \label{sec:LIM}

Linear Inverse Models are a computationally efficient, widely applied, and skillful method for predicting climate fields \citep[e.g.][]{penland1993prediction, penland1995optimal, newmanEmpiricalBenchmarkDecadal2013, perkinsLinearInverseModeling2020, meng2024reconstructing}. A LIM captures linear dynamics of anomalies about a mean state:
\begin{equation}
    \frac{d \mathbf{x}}{dt} = \bm{\mathsf{L}} \mathbf{x} + \boldsymbol{\xi},
    \label{eqn:LIM}
\end{equation}
\noindent where $\mathbf{x}$ is the state vector. $\bm{\mathsf{L}}$ is a matrix representing deterministic dynamics, and $\boldsymbol{\xi}$ is a random noise vector, which is temporally uncorrelated, but may have correlations in the state space $\mathbf{x}$. For a stable linear system, the real part of the eigenvalues of $\bm{\mathsf{L}}$ are negative \citep{penland1993prediction}. 

In this study, $\mathbf{x}$ represents low-dimensional principal components (PCs) derived from a truncated set of empirical orthogonal functions (EOFs) \citep[e.g.,][]{meng2023sacpy}. EOF truncation is applied to individual variables of interest, including 2-meter air temperature (TAS), sea-surface temperature (TOS), ocean heat content from 300 meters to the surface (OHC300), Northern Hemisphere (NH) sea-ice thickness (SIT), and NH sea-ice concentration (SIC) using SACPY \citep{meng2023sacpy,meng2024pacific}. We do not include Southern Hemisphere (SH) sea-ice due to the sparseness of SH paleoclimate proxies, and known challenges in reconciling climate model simulations of SH sea-ice with observations \citep{roach2020antarctic}. This selection of variables is guided by two primary considerations: (1) we limit the number of PCs within the LIM to prevent over-fitting, which could degrade the quality of reconstructions; (2) we exclude high-frequency atmospheric variables such as sea-level pressure (SLP) to avoid reducing the forecast skill of the primary variables of interest. We include both TAS and TOS in the LIM state vector to ensure that important variability over both land (TAS) and ocean (TOS) regions is captured, as demonstrated in Supplementary Fig. S1. This choice is necessary because we assimilate temperature-sensitive proxies located over both land and ocean. For all variables except OHC300, we select the first 15 PCs, which account for around 80\% of the total variance of each variable. Following \citet{perkinsLinearInverseModeling2020}, we select 30 PCs for OHC300 to better capture the extended memory of this variable within the LIM. Thus, the state vector is defined as:
\begin{equation}
    \mathbf{x} = [\textbf{PC}^{\rm T}_{\text{TAS}}, \textbf{PC}^{\rm T}_{\text{TOS}}, \textbf{PC}^{\rm T}_{\text{OHC300}},\textbf{PC}^{\rm T}_{\text{SIT}},\textbf{PC}^{\rm T}_{\text{SIC}}]^{\rm T}. 
    \label{eq:PC_con}
\end{equation}

\paragraph*{Linear Inverse Model training process.} We utilize output from two models in the Coupled Model Intercomparison Project Phase 5 (CMIP5) Last Millennium experiments, specifically CCSM4 and MPI-ESM-R \citep{taylor2012overview}. We choose these models primarily to maintain consistency with LMR v1 \citep{hakim2016last}, LMR v2 \citep{tardif2019last}, and LMR online \citep{perkins2021coupled}. Furthermore, seasonal climate variability statistics have not changed significantly from CMIP5 to CMIP6 \citep{brown2020comparison}. The LIM is trained separately on data from CCSM4 and MPI-ESM-R in order to capture the internal covariability and physical relationships specific to each parent model. We define four seasons by three-month averages: March-April-May (MAM), June-July-August (JJA), September-October-November (SON) and December-January-February (DJF). Prior to taking the seasonal average, model output data are placed on a 2$^{\circ}$x2$^{\circ}$ latitude-longitude grid using linear interpolation in the Climate Data Operators package \citep{schulzweida2019cdo}, and the last millennium trend for each month is removed by simple linear regression. 

EOF analysis on area-weighted variables yields the first 15 PCs for each variable (30 PCs for OHC300). Then, $\bm{\mathsf{L}}$ is calculated by
\begin{equation}
    \bm{\mathsf{L}} = \tau^{-1} \ln{\bm{\mathsf{C}}(\tau) \bm{\mathsf{C}} (0)^{-1}}.
    \label{eq:dxdt}
\end{equation}
\noindent Here $\bm{\mathsf{C}}(\tau)$ is the $\tau$-lag covariance matrix of $\mathbf{x}$, $\bm{\mathsf{C}}(\tau) = \langle\mathbf{x} (\tau) \mathbf{x}^{\rm T}(0)\rangle$, where ``$\langle\rangle$" represents a sample average. Here $\tau$ is 3 months for our seasonal LIM. The stochastic part of the dynamics, $\boldsymbol{\xi}$, has covariance matrix $\bm{\mathsf{Q}}$, such that $\langle\boldsymbol{\xi} \boldsymbol{\xi}^{\rm T}\rangle = \bm{\mathsf{Q}} $. $\bm{\mathsf{Q}}$ is calculated based on stationary statistics:
\begin{equation}
    \frac{d \bm{\mathsf{C}}(0)}{dt} = \bm{\mathsf{L}} \bm{\mathsf{C}} (0) +  \bm{\mathsf{C}} (0) \bm{\mathsf{L}}^{\rm T} + \bm{\mathsf{Q}} = 0.
    \label{eq:dcdt0}
\end{equation}

Using $\bm{\mathsf{Q}}$ and $\bm{\mathsf{L}}$, stochastic integration of (\ref{eqn:LIM}) yields a sample trajectory using the two-step integration process of \citet{penland1994balance}:
\begin{align}
    \mathbf{x}_{t+\delta t} = (\bm{\mathsf{L}} \delta t + \bm{\mathsf{I}})\mathbf{x}_{t} + \hat{\bm{\mathsf{{Q}}}} \sqrt{\mathbf{\Lambda} \delta t} \boldsymbol{\alpha} \\
    \mathbf{x}_{t + \delta t / 2} = \frac{1}{2}  (\mathbf{x}_{t+\delta t} +  \mathbf{x}_{t}),
\end{align}
\noindent where $\delta t$ is the integration time step, set at 6 hours for this study. $\bm{\mathsf{I}}$ is the identity matrix, $\hat{\bm{\mathsf{{Q}}}}$ denotes the matrix where columns are eigenvectors of $\bm{\mathsf{{Q}}}$ and $\mathbf{\Lambda}$ is the diagonal matrix of eigenvalues of $\bm{\mathsf{{Q}}}$. $\boldsymbol{\alpha}$ is a vector of independent standard normal random variables. We exclude eigenvectors associated with negative eigenvalues in $\bm{\mathsf{{Q}}}$ and normalize the remaining eigenvalues to preserve total variance, following \citet{penland1994balance} and \citet{perkinsLinearInverseModeling2020}. A complete derivation of the LIM is provided in the Supplementary Appendix. After training the LIM on output from CCSM4 and MPI-ESM-R last millennium simulations (850 C.E.--1850 C.E.) \citep{taylor2012overview}, the LIM demonstrates predictive skill to at least 12 months as evidenced by out-of-sample tests shown in Supplementary Fig. S2 and S3.

\subsection{Proxy System Models (PSM)} \label{sec:PSM}

We use the temperature-sensitive PAGES2k V2 dataset \citep{pages2k2017global} as the observational inputs for our data assimilation process. The PAGES2k V2 dataset comprises approximately 700 proxy records, mainly from tree rings, corals and ice cores. We calibrate PSMs for each proxy record, using only surface temperatures from GISTEMP v4 \citep{lenssen2019improvements} for terrestrial proxies and Sea Surface Temperature (SST) from ERSST v5 \citep{huang2017extended} for marine records. Precipitation and other moisture variables are not included in the PSM calibration. Comparisons with calibration on other instrumental observations, such as Berkeley Earth \citep{rohde2020berkeley} and MLOST \citep{smith2008improvements}, yield similar results (not shown). 

The truncated EOF basis of the LIM does not fully resolve local details of climate fields such as surface temperature and SST, which is a ``representativeness" error we account for by calibrating the PSMs in the EOF-truncated space, $\hat{\mathbf{x}} = \bm{\mathsf{U}} \bm{\mathsf{U}}^{\rm T} \mathbf{x}$, where $\bm{\mathsf{U}}^{\rm T}$ is the matrix with the first 15 PCs derived from EOF analysis: 
\begin{equation}
    \mathbf{y} = \bm{\mathsf{H}} \hat{\mathbf{x}} + \boldsymbol{\epsilon}. 
    \label{eq:PSM}
\end{equation}
\noindent Here, $\bm{\mathsf{H}}$ is the matrix that maps the climate variables to the proxy data, $\mathbf{y}$, and $\boldsymbol{\epsilon}$ is the error term.

As will be discussed in section \ref{sec:enkf}, an important factor for data assimilation is the observation error covariance matrix $\bm{\mathsf{R}} = \langle\boldsymbol{\epsilon} \boldsymbol{\epsilon}^{\rm T}\rangle$. \citet{hakim2022subseasonal} employ linear regression to estimate $\bm{\mathsf{H}}$ and a full covariance matrix $\bm{\mathsf{R}}$. In our study, we adopt a similar method but with a diagonal $\bm{\mathsf{R}}$, implying zero error covariance among errors for the proxy PSMs. The use of a full $\bm{\mathsf{R}}$ matrix is impractical because the calibration period does not provide sufficiently long overlap between each of the proxies to estimate the off-diagonal terms, and tests suggest that the diagonal values of $\bm{\mathsf{R}}$ are several orders of magnitude larger than the off-diagonal elements (not shown).

\begin{figure*}[tp]
    \centering
    \includegraphics[width=\textwidth,height=0.78\textheight,keepaspectratio]{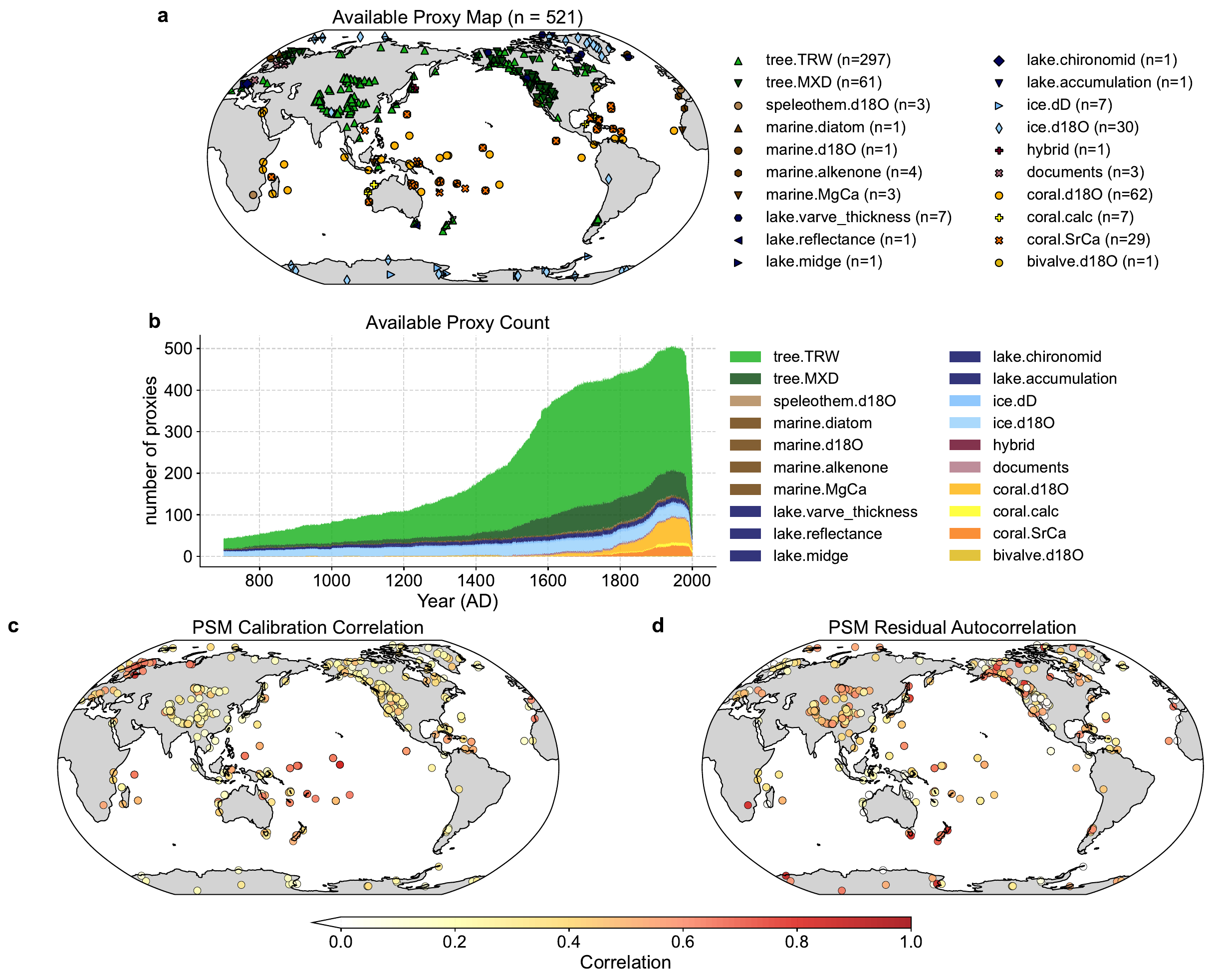}
    \caption{\textbf{Proxies from PAGES2k V2 \citep{pages2k2017global}.} \textbf{a}. Locations and counts of proxy types after filtering by the specified standards indicated by the Subsection \ref{sec:PSM}. \textbf{b}. Evolution of the number of proxies over time. \textbf{c--d}. Spatial distribution of PSM calibration correlations and 1-year lag residual (error) auto-correlations.}
    \label{fig:pages2k_used}
\end{figure*}

\paragraph*{Seasonality.} 
Seasonality refers to the specific season that a proxy's temperature represents. \citet{tardif2019last} assessed both expert-based seasonality, derived from PAGES2k metadata, and objectively-determined seasonality, given by the best correlation with instrumental data during PSM calibration. We evaluate both definitions of seasonality and find no significant differences in calibration results (compare Fig. \ref{fig:pages2k_used} with Supplementary Fig. S4), or in the PDA results as measured by instrumental verification (compare Fig.~\ref{fig:TAS_Acc_ANN} with Supplementary Fig. S5; and Fig.~\ref{fig:TAS_Acc_season} with Supplementary Fig. S6) and independent proxies  (compare Fig.~\ref{fig:Proxy_verify} with Supplementary Fig. S15; see Section~\ref{sec:verification}). We therefore use the objectively-determined seasonality to maintain consistency with LMR v2 \citep{tardif2019last} and LMR Online \citep{perkins2021coupled}. We note that sub-seasonal coral records are averaged to seasonal resolution (DJF, MAM, JJA, SON) for PSM calibration and assimilation. 

We remove proxies from assimilation if they have an insignificant correlation with local temperature or high temporal error autocorrelation. Specifically, we remove proxies that have a PSM calibration correlation below 0.05, or a one-year lag auto-correlation in PSM calibration residuals exceeding 0.90. As shown in Fig.~\ref{fig:pages2k_used}, some Pacific corals have high PSM calibration correlation but also a high 1-year-lag error auto-correlation. High error autocorrelation is problematic for Kalman filters, which assume that observation errors are uncorrelated in time.

\subsection{Ensemble Kalman Filter and Update Strategy} \label{sec:enkf}

Here we introduce the update strategy for our seasonal reconstruction, which is based on an ensemble implementation of the Kalman filter \citep[e.g.,][]{evensen2003ensemble}. 

\paragraph{Ensemble Kalman Filter.} The EnKF is extensively applied in a variety of paleo-DA tasks and has consistently shown strong performance \citep{hakim2016last, franke2017monthly,perkins2021coupled, steiger2018reconstruction,zhu2022re, valler2024mode}. The primary step in the EnKF process is the ``update,"

\begin{equation}
    \mathbf{x}_a = \mathbf{x}_p + \bm{\mathsf{K}}[\mathbf{y} - \mathcal{H}(\mathbf{x}_p)],
    \label{eq:update}
\end{equation}
\noindent where $\mathbf{x}_a$ represents the posterior (``analysis") state vector, $\mathbf{x}_p$ denotes the prior state vector and $\mathcal{H}$ is the observation operator that maps to the corresponding observation vector (i.e., the PSMs). Matrix $\bm{\mathsf{K}}$,  the Kalman Gain, is defined by,
\begin{equation}
    \bm{\mathsf{K}} = \bm{\mathsf{B H}}^{\rm T}\left[\bm{\mathsf{HBH}}^{\rm T} + \bm{\mathsf{R}}\right]^{-1},
    \label{eq:gain}
\end{equation}
\noindent where $\bm{\mathsf{B}}$ is the prior covariance matrix and $\bm{\mathsf{H}}$ is the linearization of $\mathcal{H}$. $\bm{\mathsf{R}}$ is the observation error covariance matrix derived from Equation \ref{eq:PSM}. Given that all PSMs in this study are linear, $\mathcal{H} = \bm{\mathsf{H}}$. To solve \ref{eq:update} and \ref{eq:gain} using ensemble sampling, we employ the Ensemble Square Root Filter (EnSRF) method \citep{whitaker2002ensemble} incorporating a serial observation update strategy. For the $k$th proxy, whose value is $y_k$, the update proceeds by separating the ensemble into ensemble mean ($\overline{\mathbf{x}}$) and perturbations ($\mathbf{x}^{\prime}_i$):
\begin{equation}
    \mathbf{x} = \overline{\mathbf{x}} + \mathbf{x}^{\prime}_i.
    \label{eq:xmxp}
\end{equation}

For the ensemble mean $\overline{\mathbf{x}}$, the update equation is
\begin{equation}
    \overline{\mathbf{x}_a} = \overline{\mathbf{x}_p} + \frac{\operatorname{cov}(\mathbf{x}_p, y_{e,k})}{\operatorname{var}(y_{e,k}) + R_k} (y_k - \overline{y_{e,k}}),
\end{equation}
\noindent where $y_{e,k}$ denotes the $k$th proxy estimate from from the ensemble, represented as $y_{e,k} = \overline{y_{e,k}} + y_{e,k}^{\prime}$, and $R_k$ is the $k$th proxy error variance. The ``$\operatorname{var}$" and ``$\operatorname{cov}$" operators denote the variance and covariance, respectively. Ensemble perturbations $\mathbf{x}^{\prime}_i$, are update by 
\begin{equation}
    \mathbf{x}_a^{\prime} = \mathbf{x}_p^{\prime} - \left[1 + \sqrt{\frac{R_k}{\operatorname{var} (y_{e,k}) + R_k}} \right]^{-1} \frac{\operatorname{cov}(\mathbf{x}_p, y_{e,k})}{\operatorname{var}(y_{e,k}) + R_k} (y_{e,k}^{\prime}).
\end{equation}
This ensemble update is completed for the $i$-th member using (\ref{eq:xmxp}) to obtain the full analysis state. In this study, the ensemble size is 800, which allows us to avoid ensemble inflation and localization methods \citep[e.g.,][]{anderson2012localization}. Localization techniques are complicated by the EOF state space of the LIM, so we use an ensemble large enough to minimize the need for such localization. 

\begin{figure*}[tp]
    \centering
    \includegraphics[width=\textwidth,height=0.78\textheight,keepaspectratio]{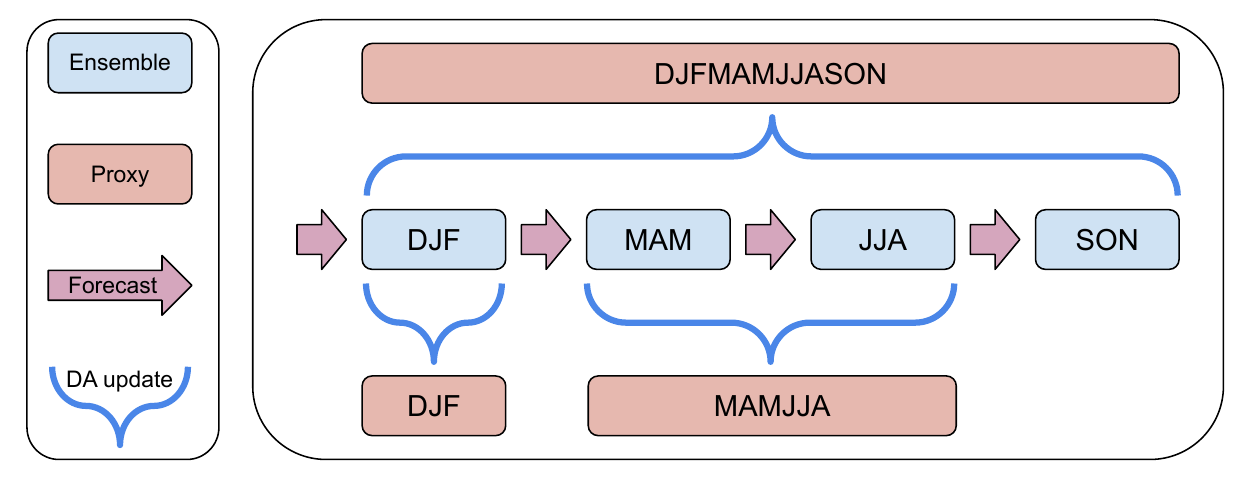}
    \caption{\textbf{LMR Seasonal update strategy.} The light blue box represents the ensemble, the rose box the proxy, and the pink arrow the forecast step from the LIM. Curly brackets denote the update from the EnKF to integrate the proxy data into updating the prior ensemble. The text within the box indicates the seasonality of either the ensemble or the proxies.}
    \label{fig:seasonalDACycle}
\end{figure*}

\paragraph{Seasonal Update Strategy} \label{para:seaonalupdatestrate}
Unlike previous PDA reconstructions \citep[e.g.][]{hakim2016last,steiger2018reconstruction,tardif2019last,perkins2021coupled} that use seasonal proxies to update the annual mean, our approach updates specific seasons corresponding to the proxy seasonality.
An example illustration of this update strategy is shown in Fig. \ref{fig:seasonalDACycle} for three proxies having different seasonality: DJF, MAMJJA and DJFMAMJJASON. When the LIM forecast completes the DJF season, the DJF proxy is used to update the DJF prior ensemble. Subsequently, the LIM advances by updating the state to the MAM and JJA seasons. Upon reaching JJA, the MAMJJA proxy is used to update MAM and JJA ensembles. Finally, when the LIM ensemble progresses to the SON season, the DJFMAMJJASON proxy is used to update the DJF, MAM, JJA and SON ensembles. In summary, our methodology emphasizes a season-to-season update mechanism. This novel update strategy has a significant impact on reconstructions of the differences between the Medieval Climate Anomaly (MCA) and Little Ice Age (LIA), as discussed in Section \ref{sec:MCALIA}. 

\subsection{Verification Metrics and Significance Test} 

We validate the LMR Seasonal reconstruction against both calibration and reanalysis datasets (discussed below) using two primary verification metrics: correlation, 
\begin{equation}
    \text{corr} = \frac{1}{n} \sum_{i=1}^n \frac{(x_i - \overline{x})(v_i - \overline{v})}{\sigma_x \sigma_v},
\end{equation}
\noindent and the coefficient of efficiency (CE) \citep{nash1970river},
\begin{equation}
    \text{CE} = 1 - \frac{\sum_{i=1}^n (v_i - x_i)^2}{\sum_{i=1}^n (v_i - \overline{v})^2}.
\end{equation}
\noindent Here an overbar ($\overline{x}$) represents a mean value, $\sigma$ represents the standard deviation, $n$ is the number of temporal samples, and $x$ and $v$ represent the reconstructed and verification values, respectively. Correlation measures errors in signal timing, whereas CE measures errors in signal timing and amplitude.

To assess whether the correlation skill of LMR Seasonal is significantly different from that of other datasets, we employ the Williams test \citep{williams1959comparison}. This test is specifically designed to compare two correlation coefficients that are statistically dependent because they share a common variable, and it accounts for this dependency using the correlation between the two predictors. Specifically, let $X$ represent our reconstruction, $Y$ represent another PDA product, and $Z$ the instrumental verification dataset. The goal is to test whether the correlation between $X$ and $Z$ ($r_{xz}$) is significantly different from that between $Y$ and $Z$ ($r_{yz}$), accounting for the correlation between $X$ and $Y$ ($r_{xy}$). The Williams test statistic is calculated as
\begin{equation}
\resizebox{0.90\linewidth}{!}{$\displaystyle t = {(r_{xz} - r_{yz})}/{\sqrt{
\frac{(1 - r_{xz}^2 - r_{yz}^2 - r_{xy}^2 + 2r_{xz}r_{yz}r_{xy})(1 + r_{xy})}
{2(n - 3)}
}},$}
\end{equation}
\noindent where $n$ is the sample size. Under the null hypothesis that $r_{xz} = r_{yz}$, the test statistic approximately follows a $t$ distribution with $(n - 3)$ degrees of freedom.

\subsection{Comparisons to other PDA Last Millennium Reconstructions} 

To assess agreement with other PDA products, we compare our results to three DA products over the last millennium: PHYDA \citep{steiger2018reconstruction}, LMR v2 \citep{tardif2019last} and LMR online \citep{perkins2021coupled}. Details on the DA methods, proxy number and time resolution for each product are given in Table \ref{tab:DAproducts}. Besides LMR-Seasonal, only PHYDA provides reconstructions for DJF and JJA (but not for SON or MAM), while LMR v2 and LMR Online are limited to  annual means. A significant distinction between the ``online" and ``offline" DA methods involves whether the prior is derived from random (time-independent) draws from an existing climate model simulation (offline) or from a forecast of the analysis at the previous assimilation time (online). An advantage of the online DA method is that the ``memory'' of past proxy information is carried to the next assimilation time. This feature is particularly vital for our seasonal DA approach, as NH trees are the dominant source of the climate signal, and are primarily sensitive to summer growing conditions. With ``online'' DA, the JJA posterior, for example, serves as the initial condition for the SON prior, informing the SON and future season's reconstructions. This results in more information persisting into seasons with fewer proxies (e.g., Winter and Spring).

\begin{table*}[t]
\centering
\begin{threeparttable}
\footnotesize
\caption{Summary of Data Assimilation Products over the Last Millennium}
\label{tab:DAproducts}
\begin{tabularx}{\textwidth}{Xcccc}
\toprule
\textbf{Name} & \textbf{PDA Method} & $N_{\text{proxy}}$ \tnote{1} & \textbf{Time Resolution}  &\textbf{Reference} \\
\midrule
PHYDA & Offline & 2978 & (Sub)Annual\tnote{2} & \cite{steiger2018reconstruction} \\
LMR v2 & Offline & 2250 & Annual & \cite{tardif2019last} \\
LMR Online & Online & 545 & Annual & \cite{perkins2021coupled} \\
LMR Seasonal& Online & 521 & Seasonal\tnote{3} & this study \\

\bottomrule
\end{tabularx}
\begin{tablenotes}
\item[1] $N_{\text{proxy}}$ denotes the total number of proxies for data assimilation.
\item[2] ``(Sub)Annual" refers to PHYDA's time resolutions finer than a year, because it has the annual mean, DJF and JJA reconstructions, and monthly Niño 3.4 Index reconstructions.
\item[3] ``Seasonal" refers to 4 time steps (MAM, JJA, SON, DJF) every year.
\end{tablenotes}
\end{threeparttable}
\end{table*}

\section{Verification} \label{sec:verification}

We verify our reconstruction using instrumental observations and proxy data, as described below. For instrumental verification, we use 2m air temperature from HadCRUT5 \citep{morice2021updated} and the ERA-20C Reanalysis \citep{poli2016era}, ocean temperature data from Hadley EN4 \citep{good2013en4}, SST from HadISST \citep{rayner2003global}, sea-ice concentration from the satellite dataset of \citet{fetterer2017sea}, and NH sea-ice concentration and sea-ice thickness from PIOMAS20C \citep{schweigerArcticSeaIce2019}. For proxy data verification we use the PAGES2k V2 dataset \citep{pages2k2017global} by withholding some proxies from assimilation using the bootstrap procedure described below.

\subsection{Instrumental Verification}

\begin{figure*}[tp]
    \centering
    \includegraphics[width=\textwidth,height=0.78\textheight,keepaspectratio]{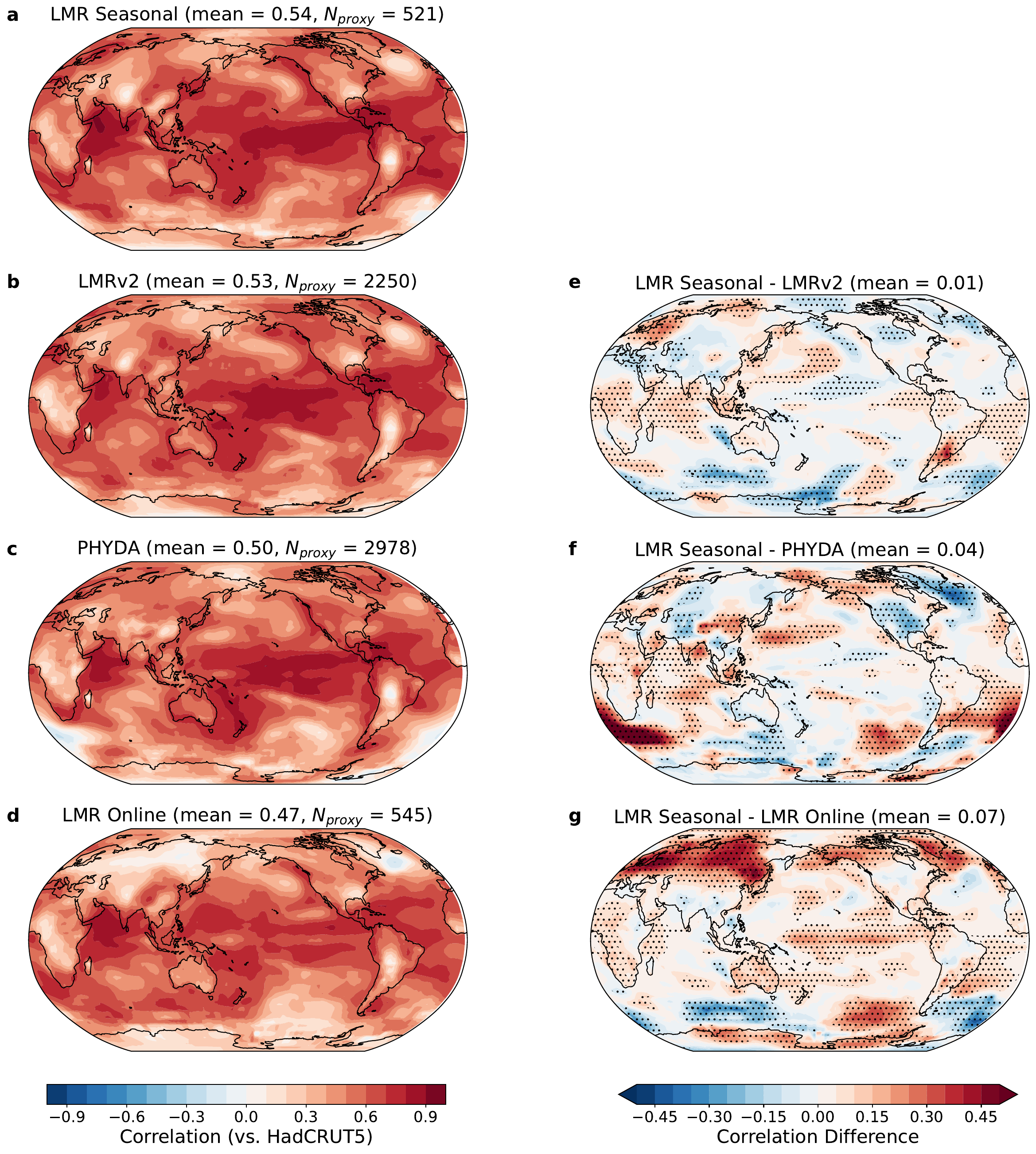}
    \caption{\textbf{Annual mean surface temperature instrumental verification.} \textbf{a--d.} Correlation between various DA reconstructions ensemble mean and HadCRUT5 \citep{morice2021updated} 2m air temperature during 1880-–2000. Results are shown for (\textbf{a} LMR Seasonal, \textbf{b} LMRv2, \textbf{c} PHYDA, \textbf{d} LMR Online) with the global-mean correlation and the number of used proxies given in the title for each subpanel. Correlation difference between LMR Seasonal (\textbf{a}) and other reconstructions are shown for (\textbf{e} LMRv2, \textbf{f} PHYDA, \textbf{g} LMR Online), with global-mean correlation differences indicated in the titles. Black dots show the correlation difference values passing the confidence level of 95\% using Williams test \citep{williams1959comparison}.}
    \label{fig:TAS_Acc_ANN}
\end{figure*}

Measured by correlation with the HadCRUT5 instrumental dataset during 1880--2000~CE, the LMR Seasonal reconstruction skill in annual-mean 2m air temperature is similar to or better than other reconstructions in the global mean (Fig.~\ref{fig:TAS_Acc_ANN}) despite assimilating fewer proxies. The spatial correlation pattern shows that LMR Seasonal performs better than LMR v2 primarily in Europe, the North Pacific Ocean, the Indian Ocean, and the South Atlantic Ocean, but less well over the Southern Ocean and portions of Asia. Compared to PHYDA, the most pronounced differences are found in the Southern Ocean and the Indian Ocean. Similar results are found when verifying against the ERA-20C Reanalysis (Supplementary Fig. S7 and S8). The limited number of proxies in the Southern Hemisphere means that the main signal for reconstructing the Southern Ocean relies heavily on long-distance teleconnections in the model prior. Since we do not use covariance localization, it is possible that bias in the Southern Ocean teleconnections from LIM forecasts degrade the performance of LMR Seasonal in this location. We note that in Antarctica, where there are abundant ice core records \cite[e.g.,][]{steig2013,stenni2017}, the reconstructions perform comparatively well, with LMR Seasonal generally being superior.   

\begin{figure*}[tp]
    \centering
    \includegraphics[width=\textwidth,height=0.78\textheight,keepaspectratio]{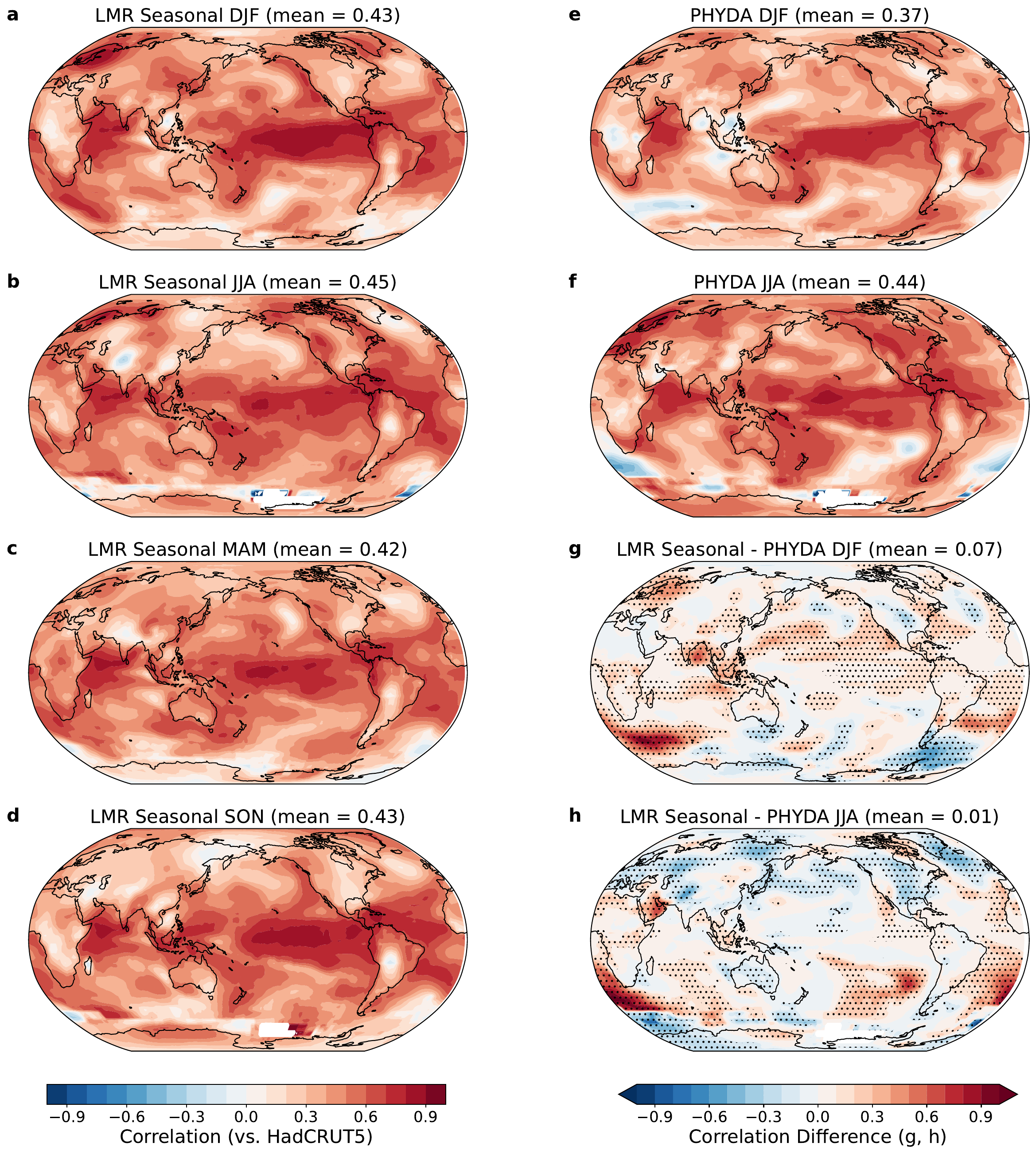}
    \caption{\textbf{Surface temperature seasonal instrumental verification}  \textbf{a--d.} The correlations between the LMR Seasonal ensemble mean and HadCRUT5 \citep{morice2021updated}. For DJF (\textbf{a}), MAM (\textbf{c}), JJA (\textbf{b}), and SON (\textbf{d}) during 1880--2000. \textbf{e--f.} Correlations between PHYDA and HadCRUT5 are shown for DJF (\textbf{e}) and JJA (\textbf{f}). \textbf{g--h.} The correlation differences between LMR Season and PHYDA in DJF (\textbf{g}) and JJA (\textbf{h}).  Black dots show the correlation difference values passing the confidence level of 95\% using Williams test \citep{williams1959comparison}.}
    \label{fig:TAS_Acc_season}
\end{figure*}

Seasonal verification against HadCRUT5 reveals that LMR Seasonal has skill globally with positive correlations almost everywhere (Fig.~\ref{fig:TAS_Acc_season}). Compared to PHYDA, which is the only other reconstruction that includes DJF and JJA reconstructions, LMR Seasonal performs relatively better during DJF than JJA, which we attribute to the sparse proxy data during DJF compared to JJA. PHYDA performs better during Northern Hemisphere summer, particularly over North America and Eurasia, which we attribute to the much large number of tree-ring proxies that PHYDA assimilates in these locations.

We also compare LMR Seasonal with ModE-RA, a monthly offline DA product \citep{valler2024mode} from 1421 to 1900, as shown in Supplementary Figs.~S9 and S10. The correlation is largest in the locations and seasons when proxy availability is highest: JJA over western North America, central Asia, and northern Europe. This shows that the information content from the proxies is similar in both reconstructions, but how that information is spread in space and time is different. We attribute this to 1) the state-dependent prior covariances in LMR Seasonal (static in ModE-RA), and 2) the way in which information is transmitted from times of high proxy availability to low (summer to winter). . After 1900, ModE-RA assimilates a large number of instrumental observations, which results in higher correlations with LMR Seasonal.

\begin{figure*}[tp]
    \centering
    \includegraphics[width=\textwidth,height=0.78\textheight,keepaspectratio]{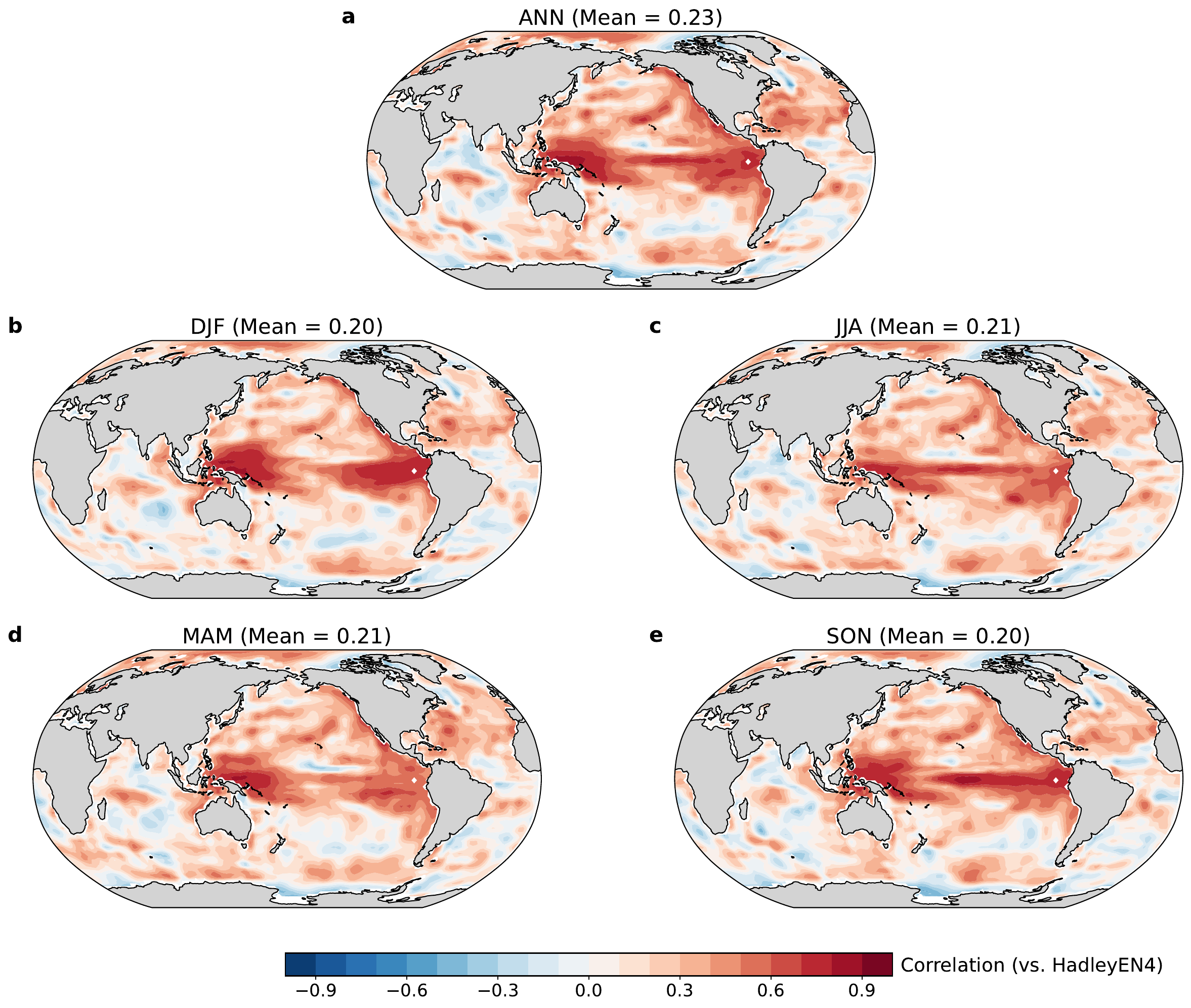}
    \caption{\textbf{Ocean heat content from 300-meter to the surface (OHC300) instrumental verification.} Correlation between LMR Seasonal OHC300 ensemble mean and HadleyEN4 OHC300 \citep{good2013en4} over the period 1940--2000 for the annual mean (\textbf{a}), DJF (\textbf{b}), JJA (\textbf{c}), MAM (\textbf{d}), and SON (\textbf{e}). Global-mean correlations are indicated in the titles.}
    \label{fig:OHC300_Acc}
\end{figure*}

\begin{figure}[tp]
    \centering
    \includegraphics[width=\linewidth,height=0.75\textheight,keepaspectratio]{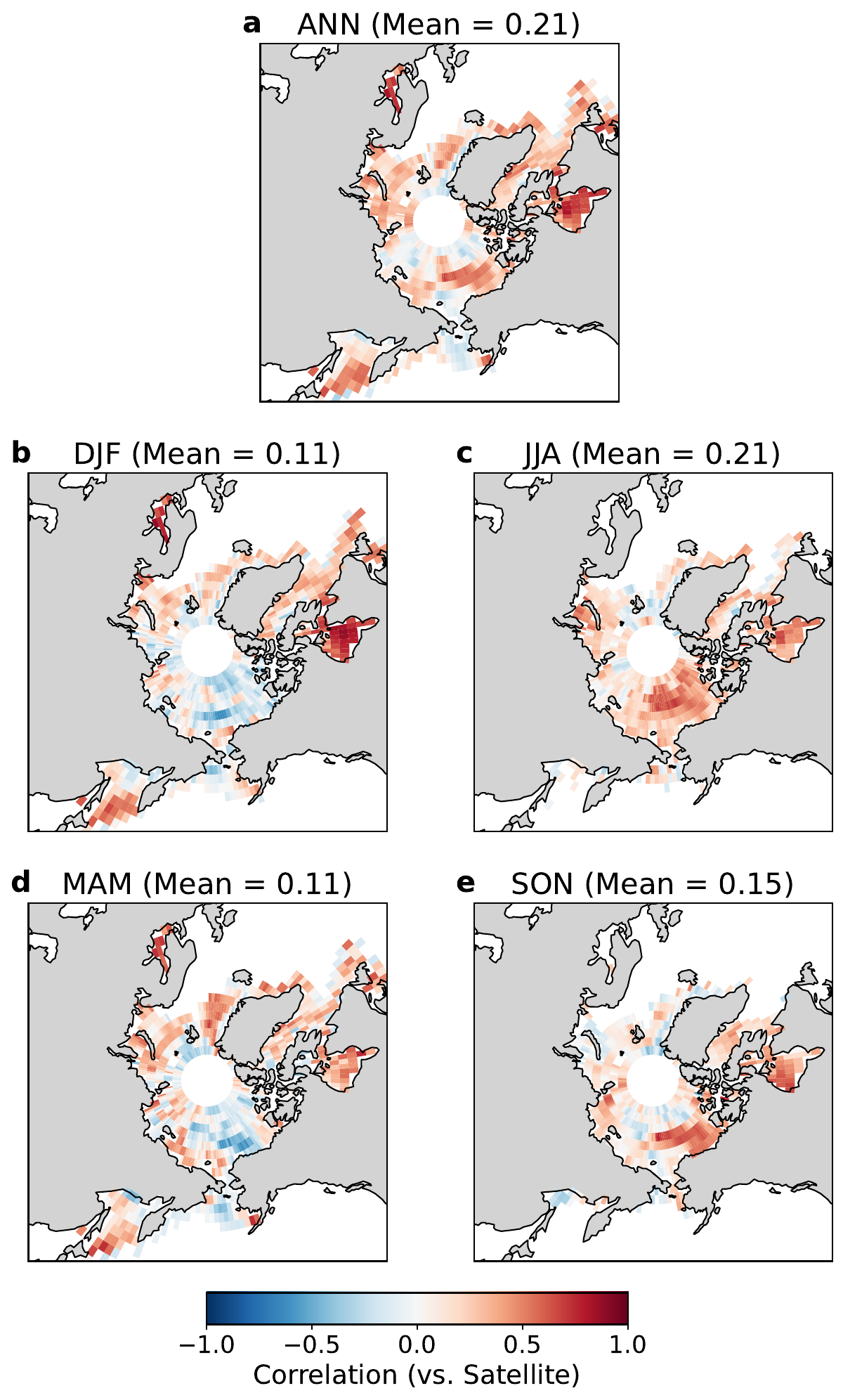}
    \caption{\textbf{Northern Hemisphere sea-ice concentration (SIC) instrumental verification.} Correlation between the LMR Seasonal SIC ensemble mean and satellite SIC data \citep{fetterer2017sea} during 1980--2000 are presented for the annual mean (\textbf{a}), DJF (\textbf{b}), JJA (\textbf{c}), MAM (\textbf{d}), and SON (\textbf{e}). Global-mean correlations are indicated in the titles.}

    \label{fig:SIC_Acc}
\end{figure}

We verify upper-300m ocean-heat content (OHC300) against the Hadley EN4 dataset \citep{good2013en4}(Fig.~\ref{fig:OHC300_Acc}) and Arctic sea-ice concentration (SIC) against the satellite observations \citet{fetterer2017sea} (Fig.~\ref{fig:SIC_Acc}). Despite not incorporating any direct observations of these quantities, the LMR Seasonal reconstruction show high correlation with the verification datasets. For OHC300, skill is highest in the tropical Pacific, the eastern North Pacific, and northern Atlantic Ocean regions. Skill in SIC is highest in Hudson Bay, and near sea-ice edges, especially around Greenland and the Barents Sea. Skill is lowest in the Beaufort Sea during DJF and MAM, seasons where proxies are least abundant. In contrast, SIC generally exhibits higher correlations in JJA compared to other seasons. We speculate that this is due to SIC having a stronger correlation with 2m air temperature in JJA compared to other seasons \citep{blanchard2011persistence}.

\begin{figure*}[tp]
    \centering
    \includegraphics[width=\textwidth,height=0.78\textheight,keepaspectratio]{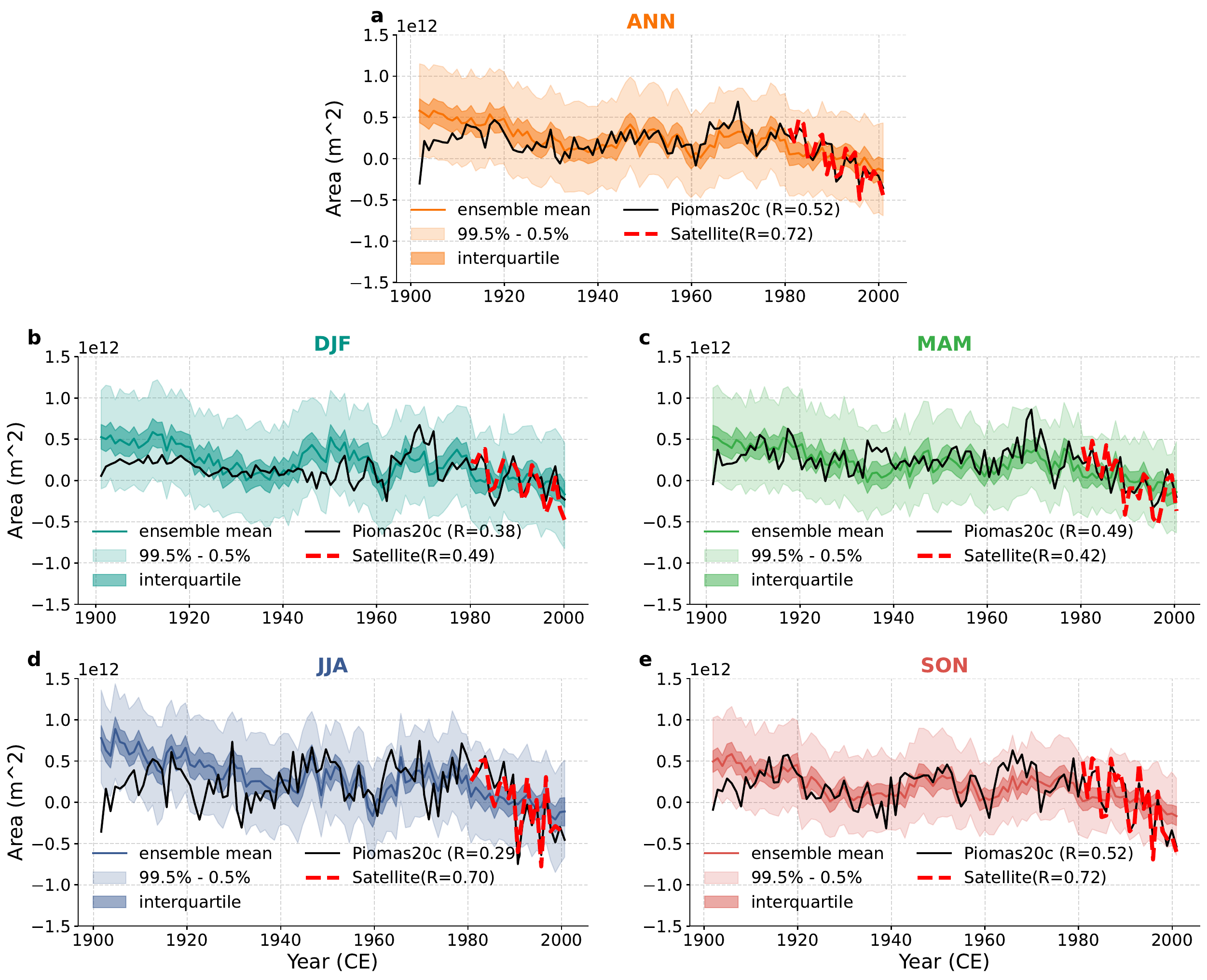}
        \caption{\textbf{Northern Hemisphere sea-ice area (SIA) verification.} Temporal comparison of the LMR Seasonal reconstructed NH sea-ice area series (colored curves) against Piomas20c \citep{schweigerArcticSeaIce2019} sea-ice area (black solid curve), and satellite  sea-ice area \citep{fetterer2017sea} (red dashed curve) in the annual mean (\textbf{a}), DJF (\textbf{b}), MAM (\textbf{c}), JJA (\textbf{d}), and SON (\textbf{e}). The reference time period for anomalies is 1980--2000. Dark shading denotes the ensemble interquartile range, and light shading the 0.5\% to 99.5\% interval. }
    \label{rfig:sia}
\end{figure*}

Furthermore, we also compare NH sea-ice area (SIA) and sea-ice volume (SIV) with PIOMAS20C \citep{schweigerArcticSeaIce2019}, which is produced by forcing a sea-ice model with the ERA20C atmospheric reanalysis (Fig.~\ref{rfig:sia}, and Supplementary Fig. S12). Discrepancies appear in the early 20th century: PIOMAS-20C shows a rapid increase in SIA around 1900, likely due to sensitivity to the initial conditions (PIOMAS20C is initialized with HadISST, version 2.1.0.0), whereas the rest of the 20th century  shows much better agreement. The more persistent quantity, SIV, shows differences through the first half of the 20th century. LMR Seasonal, which begins data assimilation in 700 CE, benefits from proxy information before 1901 and thus avoids abrupt early-century adjustments. We also compare LMR Seasonal SIA with IAPICE1 \citep{semenov2024arctic}, which reconstruction of SIC for March and September during 1901--2019 based on the statistical co-variability between sea ice concentration, sea surface temperature, and sea level pressure. LMR Seasonal shows less temporal variance in sea-ice area, which we attribute to ensemble averaging and fewer numbers of proxies relative to instrumental observations (see Supplementary Fig.~11.).  We also compare sea-ice extent (SIE) with the August reconstruction from \citet{kinnard2011reconstructed} over the last millennium (Supplementary  Fig.~14). Their reconstruction also exhibits greater temporal variability, likely due to its focus on a single month (August), in contrast to the seasonal (JJA) resolution used in LMR Seasonal, and ensemble averaging in LMR Seasonal. The correlation between the two reconstructions is 0.26.

\begin{figure*}[tp]
    \centering
    \includegraphics[width=\textwidth,height=0.78\textheight,keepaspectratio]{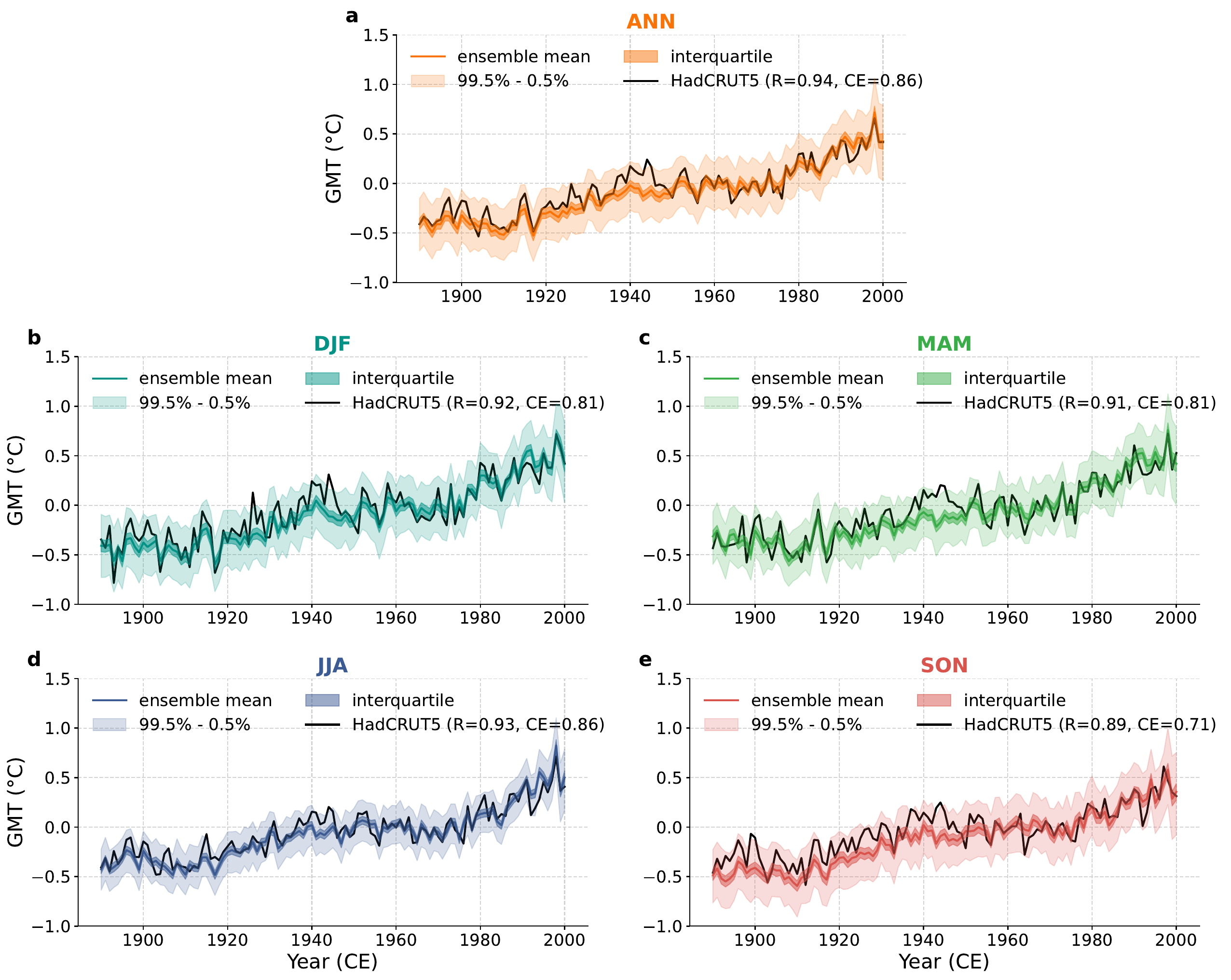}
    \caption{\textbf{Global mean surface temperature (GMT) instrumental verification.} Temporal verification of the ensemble-mean LMR Seasonal reconstructed GMT series (colored curves) against HadCRUT5 \citep{morice2021updated} GMT (black solid curve) in annual mean (\textbf{a}), DJF (\textbf{b}), MAM (\textbf{c}), JJA (\textbf{d}), and SON (\textbf{e}). The reference time period for anomalies is 1950--1980. For each reconstruction, dark shading denotes the ensemble interquartile range, and light shading the 0.5\% to 99.5\% interval. R = correlation, CE = coefficient of efficiency.}
    \label{fig:GMT}
\end{figure*}

\begin{figure*}[tp]
    \centering
    \includegraphics[width=\textwidth,height=0.78\textheight,keepaspectratio]{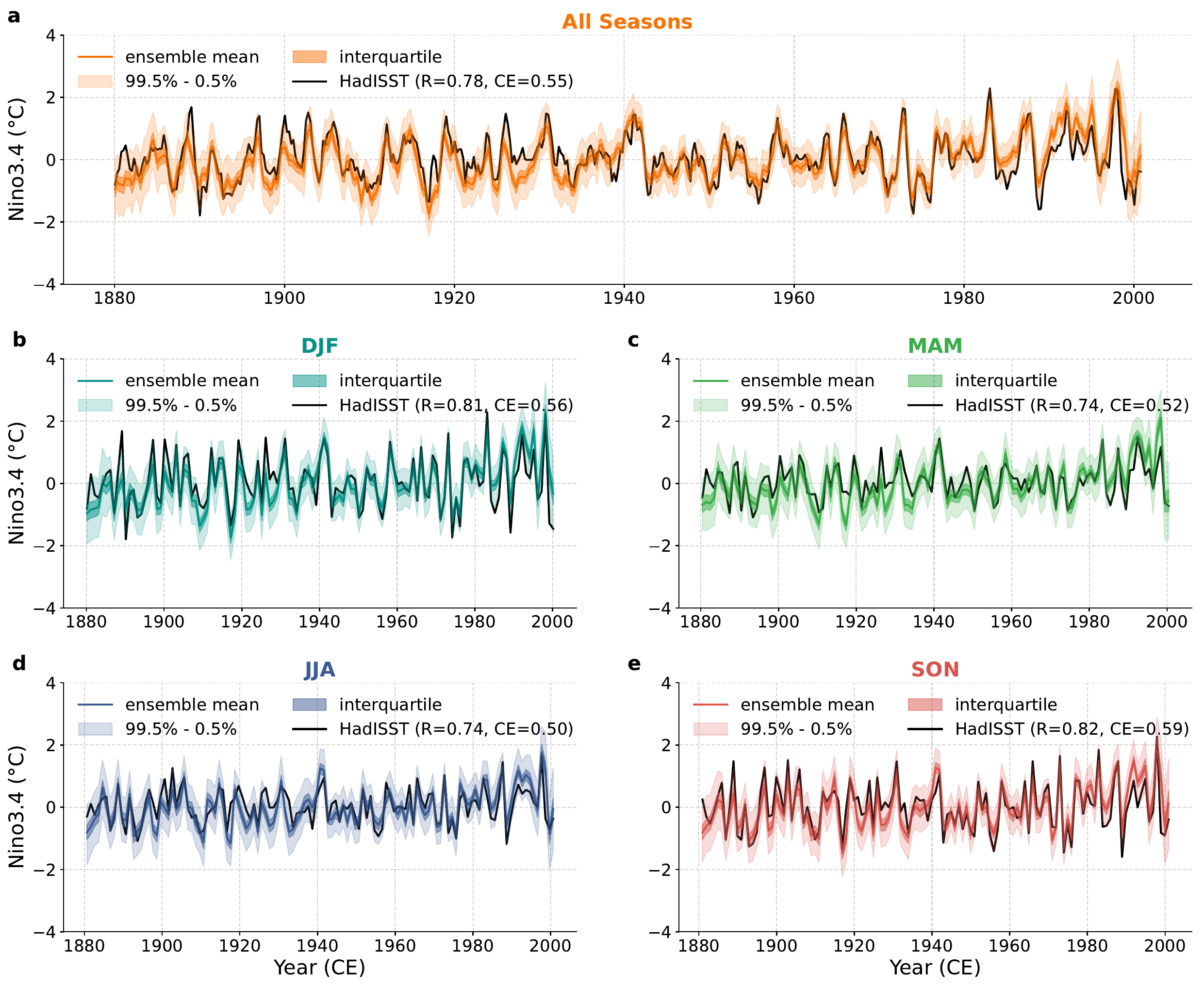}
    \caption{\textbf{Ni\~{n}o3.4 index instrumental verification.} Temporal verification of the ensemble mean LMR Seasonal reconstructed Ni\~{n}o3.4 Index (colored curves) against HadISST \citep{rayner2003global} (black solid curve) in all seasons (\textbf{a}), DJF (\textbf{b}), MAM (\textbf{c}), JJA (\textbf{d}), and SON (\textbf{e}). Dark shading denotes the interquartile range, and light shading the 0.5\% to 99.5\% interval. R = correlation, CE = coefficient of efficiency.}
    \label{fig:Nino34}
\end{figure*}

Comparing the LMR Seasonal global mean temperature (GMT) and Ni\~{n}o3.4 Index with values from the instrumental datasets HadCRUT5 and HadISST shows highly skillful reconstructions (Figs. \ref{fig:GMT} and \ref{fig:Nino34}). Specifically, GMT has a correlation with HadCRUT5 of about 0.9 in all seasons and in the annual mean. The Ni\~{n}o3.4 Index, which represents the intensity of ENSO, shows a correlation (CE) around 0.8 (0.55) in all seasons. The use of seasonal coral data significantly improves the accuracy of the reconstruction compared to annualized coral data, which shows a correlation (CE) of approximately 0.7 (0.3) with the annual coral reconstructions and HadISST (not shown). In addition, compared to PHYDA, our reconstruction has higher correlation and CE in all seasons except in JJA's correlation (Supplementary Fig. S13). The greatest contribution to improved ENSO reconstruction likely comes from the online DA scheme, since ENSO exhibits forecast skill on seasonal to annual timescales \citep{mcphaden2006enso,perkinsLinearInverseModeling2020, meng2024reconstructing, meng2024pacific}; information from the previous season persists to subsequent seasons, providing a more accurate prior. 

\begin{figure*}[tp]
    \centering
    \includegraphics[width=0.85\textwidth,height=0.78\textheight,keepaspectratio]{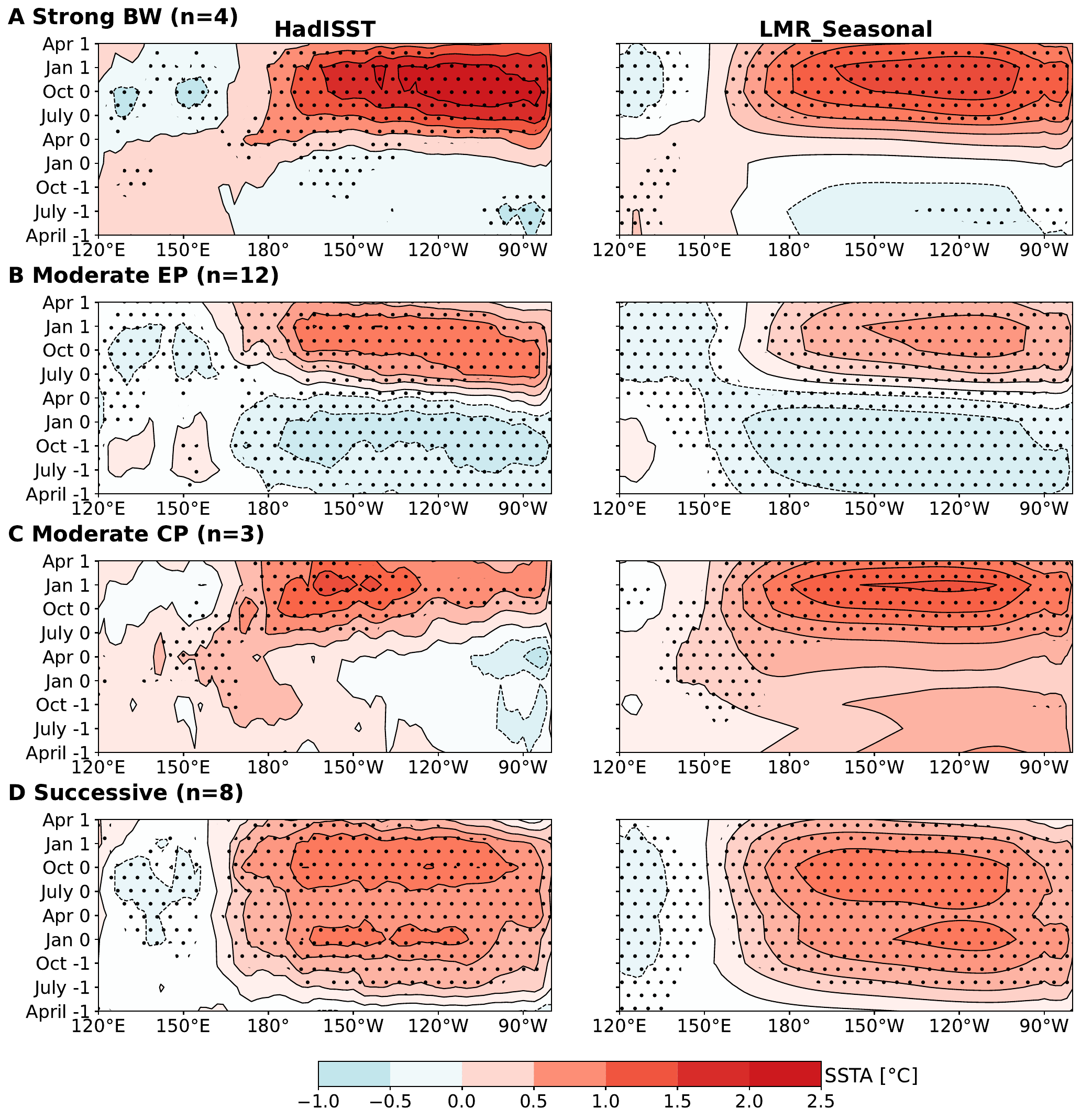}
    \caption{\textbf{Verification of the four classes of El Ni\~{n}o onset evolution of \citet{wang2019historical} during 1900--2000.} The left column displays composite analyses from HadISST \citep{rayner2003global}, and the right column shows the LMR Seasonal Reconstruction ensemble mean. Rows show composite averages of the four El Ni\~{n}o classes: Strong Basin-Wide (SBW) (\textbf{A}), Moderate Eastern Pacific (MEP) (\textbf{B}), Moderate Central Pacific (MCP) (\textbf{C}), and Successive (\textbf{D}). Black dots show the composite SST anomaly values passing the confidence level of 95\% using the Student T-test.}
    \label{fig:en_evolutions}
\end{figure*}

\begin{figure*}[tp]
    \centering
    \includegraphics[width=\textwidth,height=0.78\textheight,keepaspectratio]{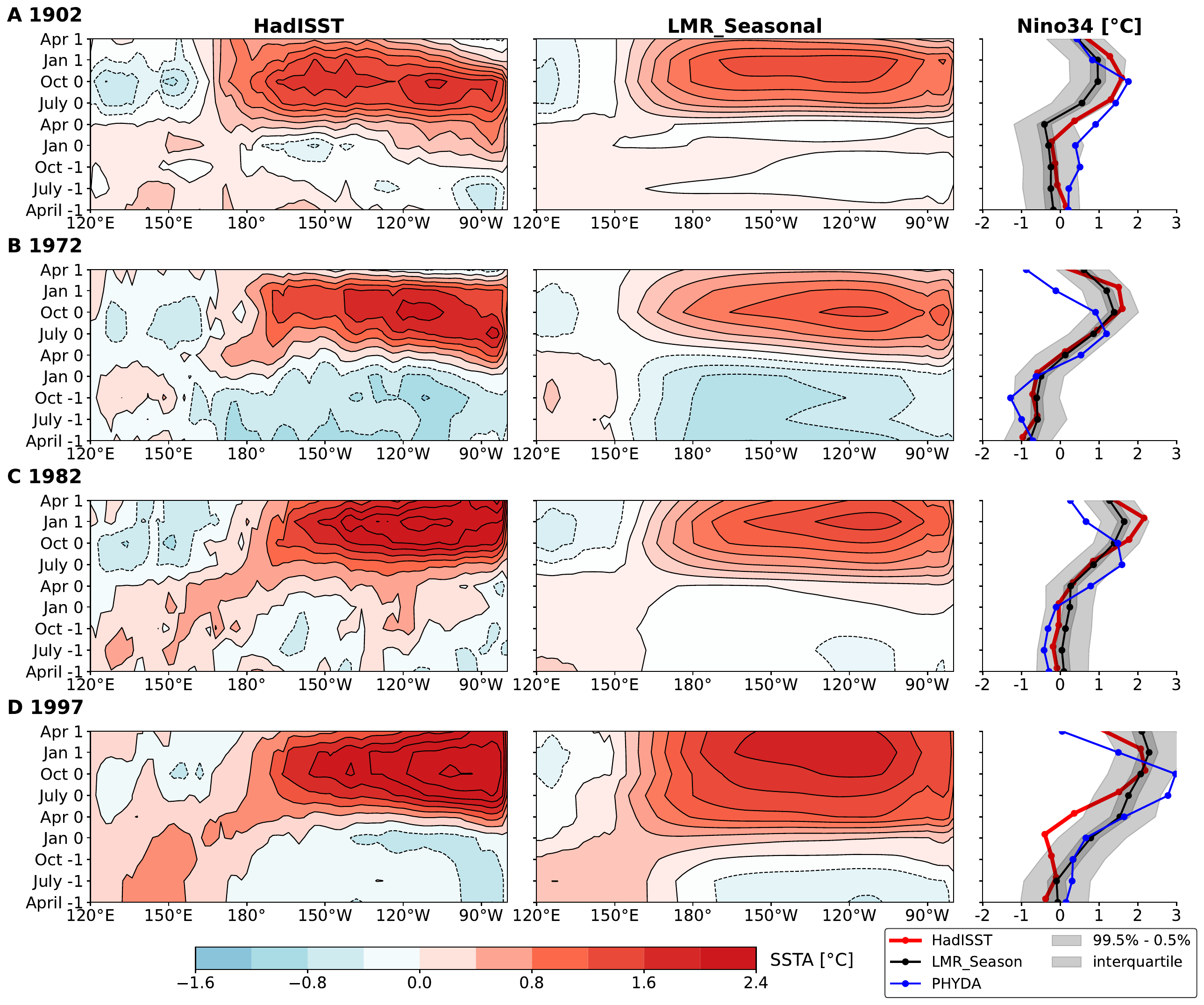}
    \caption{\textbf{Verification of four Strong Basin-Wide (Super) El Ni\~{n}o Cases' onset evolution (1902, 1972, 1982, and 1997).} The left column shows the evolution in HadISST \citep{rayner2003global}, and the middle column represents the LMR Seasonal Reconstruction ensemble mean. The right column depicts the time series of the Ni\~{n}o 3.4 Index from HadISST (red), PHYDA (blue) and LMR Seasonal reconstruction mean (black). The dark shading denotes the interquartile range, and light shading denotes the central 99\% confidence interval.}
    \label{fig:sp-en}
\end{figure*}

The ENSO reconstruction allows us to investigate the variability of El Ni\~{n}o over the last millennium with a much larger sample than is available during the instrumental period. Following \citet{wang2019historical}, we categorize El Ni\~{n}o events into four classes based on their onset phase evolution: Strong Basin Wide (SBW), Moderate Eastern Pacific (MEP), Moderate Central Pacific (MCP), and Successive. The composite analysis of El Ni\~{n}o cases from 1900 to 2000 based on this classification is illustrated in Fig. \ref{fig:en_evolutions}. As described by \citet{wang2019historical}, SBW events are characterized by a combination of eastward SST anomalies (SSTA) from the western Pacific and westward SSTA from the eastern Pacific, leading to strong warming events (``super El Ni\~{n}o"). In contrast, MEP and MCP events are defined by westward and eastward SSTA from the eastern and western Pacific, respectively, resulting in moderate warming. Successive cases describe two consecutive years of sustained El Ni\~{n}o conditions. Compared to HadISST data, the LMR Seasonal reconstruction successfully captures most of the seasonal evolution of these four El Ni\~{n}o classes, although the amplitude of the SBW and MEP cases is smaller in the reconstruction than in HadISST. The most significant discrepancies occur for MCP due to the small sample size (3 events), since most cases are after the year 2000; many regions still align well with HadISST observations. Furthermore, as depicted in Fig. \ref{fig:sp-en}, all four SBW cases demonstrate consistent evolution with the reconstructed Ni\~{n}o3.4 index evolution, closely following the HadISST time series. In comparison, the PHYDA reconstruction does not align as closely with HadISST, especially during JJA and SON.

In summary, instrumental verification shows that the LMR Seasonal reconstruction faithfully captures a wide range of coupled atmosphere--ocean--sea-ice climate variability in space and time during the instrumental period.

\subsection{Independent Proxy Verification}

\begin{figure*}[tp]
    \centering
    \includegraphics[width=\textwidth,height=0.78\textheight,keepaspectratio]{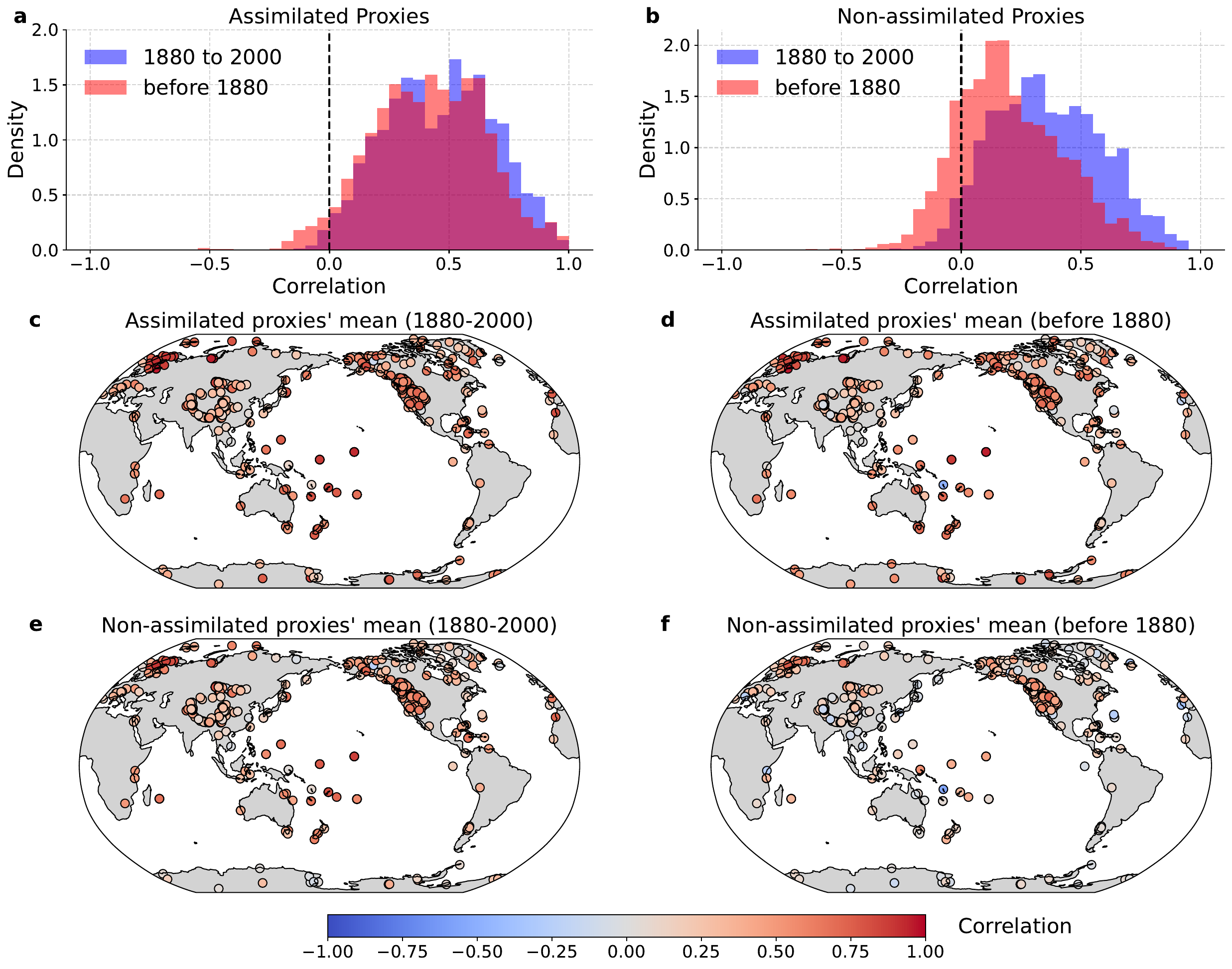}
    \caption{\textbf{Verification of LMR Seasonal against assimilated and non-assimilated proxy data.} The top row illustrates the distribution of correlation values between proxy values and LMR Seasonal estimates for assimilated (\textbf{a}) and non-assimilated (\textbf{b}) proxy data from 1880--2000 (blue) and prior to 1880 (red). The middle row shows proxy time-mean correlation maps for assimilated proxies during 1880--2000 (\textbf{c}) and before 1880 (\textbf{d}), and the bottom row for non-assimilated proxies during 1880--2000 (\textbf{e}) and before 1880 (\textbf{f}).}
    \label{fig:Proxy_verify}
\end{figure*}

To assess the robustness of LMR Seasonal in the pre-instrumental period (800--1850~CE), we validate against proxies left out of the assimilation process following \citet{hakim2016last}. We employ the bootstrap method, randomly omitting 20\% of the proxies and conducting DA across 50 epochs. For each epoch, the proxies are forward modeled from the reconstructed climate states using the PSM (\ref{eq:PSM}) for each proxy, yielding a direct comparison of the LMR Seasonal reconstruction to both the assimilated and independent proxy chronologies. The comparison is summarized by the time series correlation between the reconstructed and actual proxy time series. For the assimilated proxies, we find that the distribution of correlation values during the calibration and pre-calibration periods is very similar, suggesting a robust PSM relationship (Fig. \ref{fig:Proxy_verify}). Results for non-assimilated proxies are similar, but with somewhat lower correlation values. There are no significant differences between these results and those when seasonality is defined by the expert-based seasonality defined in the PAGES2K database results (cf. Fig. \ref{fig:Proxy_verify} and Supplementary Fig. S15). This indicates that the results are insensitive to the exact definition of proxy seasonality. 

\section{Last Millennium Seasonal Temperature Trends, Medieval Climate Anomaly and Little Ice Age} \label{sec:MCALIA}

\begin{figure*}[tp]
    \centering
    \includegraphics[width=\textwidth,height=0.78\textheight,keepaspectratio]{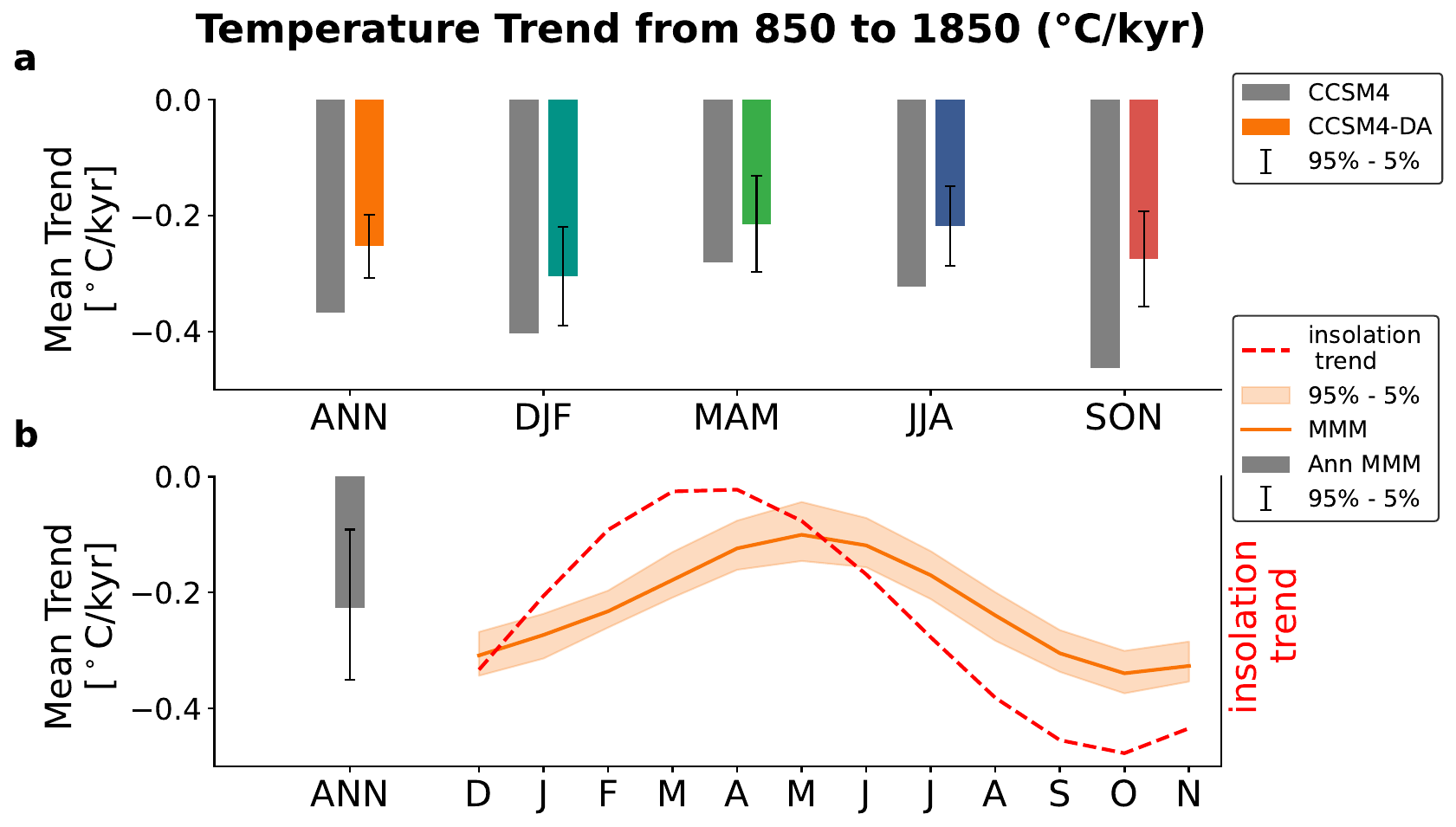}
    \caption{\textbf{Comparison of Last Millennium (LM) global-mean temperature trends for LMR Seasonal and the Climate Model LM simulations.} \textbf{a.} The global mean temperature trend for the annual mean and each season. The gray bar denotes the CCSM4 LM simulation, and the colors LMR Seasonal. Error bars indicate the 90\% ensemble confidence interval. \textbf{b.} The global-mean temperature trends for the annual mean (gray bar) and each month (orange solid line) from the Last Millennium CMIP5 Multi-Model Large Ensemble (CCSM4, CESM-LME, CSIRO-Mk3L-1–2, MPI-ESM-P, IPSL-CM5A-LR and HadCM3) \citep{taylor2012overview}. The top-of-atmosphere insolation trend (right $y$-axis) is shown in the red dashed line \citep{laskar2004long}. Error bars and orange shading represent the central 90\% confidence interval.}
    \label{fig:temp_trend}
\end{figure*}

From 850--1850~CE, most proxy evidence suggests that Earth experienced a cooling climate trend driven by orbital forcing and significant volcanic eruptions \citep{mcgregor2015robust}. We compare our reconstructed seasonal temperature trends with seasonal trends from the CCSM4 Last Millennium simulation, as depicted in Fig. \ref{fig:temp_trend} and Supplementary Fig. S16. Both the reconstructed and modeled trends show enhanced cooling in DJF and SON relative to MAM and JJA, which have smaller trends. This seasonal difference is attributed to the delayed climate response to orbital forcing (Fig.\ref{fig:temp_trend}b, red dashed curve), as discussed by \citet{lucke2021orbital}. It is important to note that the LIM trained on this model simulation has no trend, and no season-specific variability (i.e., a single LIM is used for all seasons); the reconstructed seasonal trends arise solely from assimilation of proxy data.

\begin{figure*}[tp]
    \centering
    \includegraphics[width=\textwidth,height=0.78\textheight,keepaspectratio]{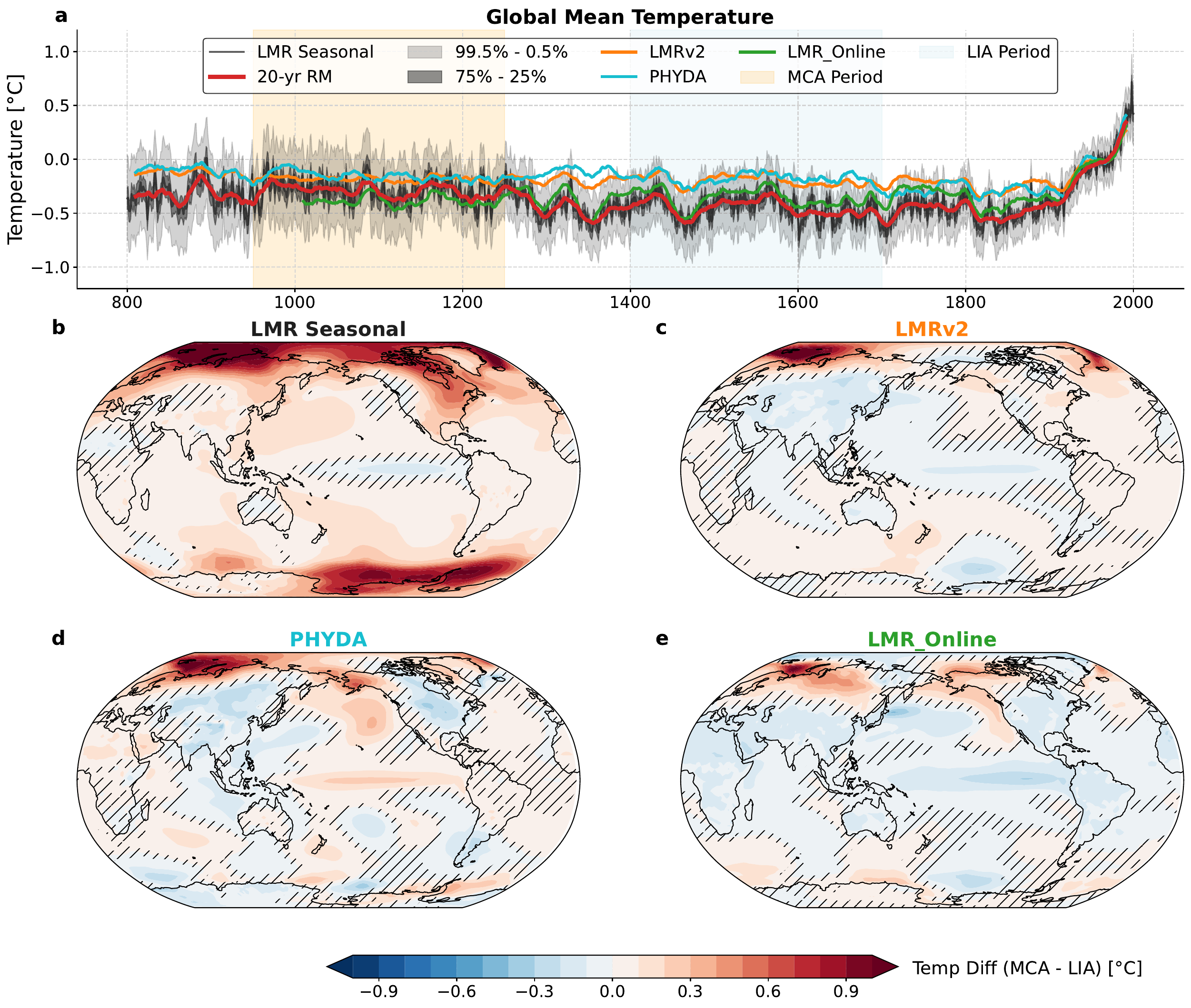}
    \caption{\textbf{Differences Between the Medieval Climate Anomaly (MCA, 950CE--1250CE) and Little Ice Age (LIA, 1400CE--1700CE) in four DA Reconstructions.} \textbf{a.} Global Mean Surface Temperature (GMT) 20-year running mean in LMR Seasonal (red), LMRv2 (yellow), LMR Online (green), and PHYDA (blue). The black solid curve represents the LMR Seasonal unsmoothed GMT, dark shading the interquartile range, and light shading the central 99\% confidence interval. \textbf{b--e.} Global temperature pattern differences between MCA and LIA from LMR Seasonal (\textbf{b}), LMRv2 (\textbf{c}), PHYDA (\textbf{d}), and LMR Online (\textbf{e}). Hatching denotes regions that do not pass the 95\% confidence level according to Student's t-test.}
    \label{fig:MCA_LIA_pattern}
\end{figure*}

Over the Last Millennium, two significant periods of multicentennial climate variability are the Medieval Climate Anomaly (MCA) and the Little Ice Age (LIA). Following \citet{mann2009global}, we define the MCA as the period from 950 CE to 1250 CE and the LIA from 1400 CE to 1700 CE. Comparing the annual-mean global-mean temperature in LMR Seasonal with three other reconstructions, we find that LMR Seasonal has larger multicentennial variability, especially with respect to the MCA-LIA difference (Fig.~\ref{fig:MCA_LIA_pattern}a). The spatial pattern of temperature differences between these two time periods (MCA-LIA) reveals a common pattern of Arctic-amplified warming among the reconstructions (Fig.~\ref{fig:MCA_LIA_pattern}b--e). Notable differences are the much larger amplitude signal in LMR Seasonal, and the opposite sign of tropical Pacific temperature difference when compared with PYHDA. Moreover, LMR Seasonal shows an MCA-LIA pattern over parts of Antarctica and the Southern Ocean that is mostly absent in the other reconstructions. This pattern is consistent with the independent temperature reconstruction of \citet{orsi2012little} based on a borehole thermometry analysis in West Antarctica (Supplementary Fig. S17). We attribute these high-latitude differences to polar amplification having larger amplitude on seasonal time scales, which leads to more signal in the annual mean (Supplementary Fig. S18).

As discussed above with regard to seasonal temperature trends, summer exhibits the least cooling trend over the last millennium, attributable to differences in insolation trends and seasonal lag due to ocean heat content \citep{lucke2021orbital}. Most proxies, especially NH tree ring width and latewood density, predominantly record JJA temperatures \citep[e.g.,][]{briffa1992tree,anchukaitis2017last}. Annual-mean data assimilation dilutes the influence of JJA proxies, which likely reduces cooling trends over the Last Millennium. Our seasonal-update strategy ensures that connections between seasons are dynamically connected by the LIM, rather than static as in offline DA approaches. We hypothesize that these factors collectively contribute to the distinct differences we observe between the MCA and LIA. To test this hypothesis, we perform another experiment, allowing seasonal proxies to update only the annual mean during assimilation. Results show that the global-mean temperature difference between the MCA and LIA decreases by 30\%, from 0.15°C to 0.10°C. In this case, our seasonal-update strategy appears to be essential to reconstructing the magnitude of the MCA-LIA difference.

\begin{figure*}[tp]
    \centering
    \includegraphics[width=\textwidth,height=0.78\textheight,keepaspectratio]{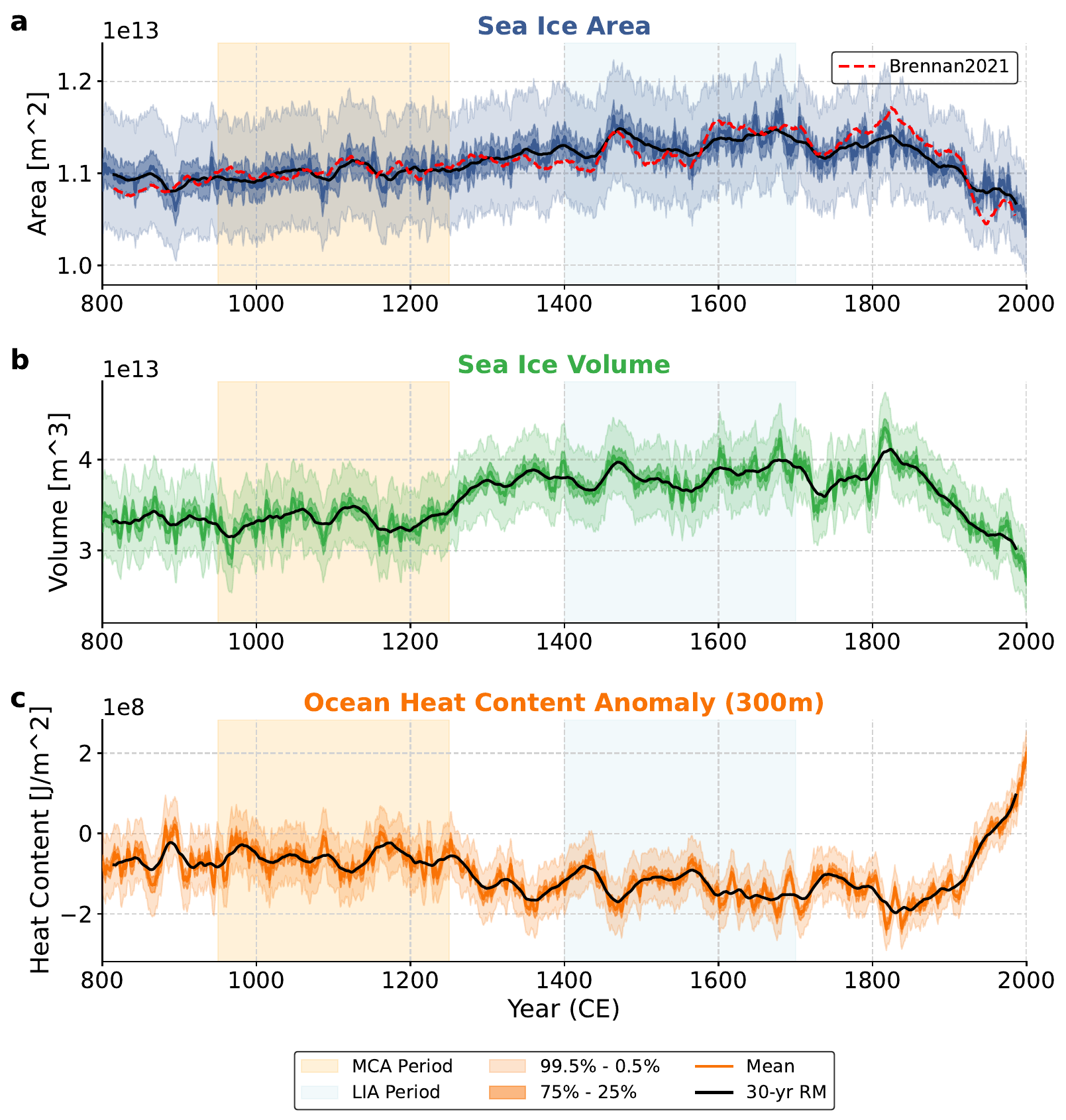}
    \caption{\textbf{Time series of Arctic sea-ice area (a) and volume (b), and upper 300m ocean heat content anomaly (c) over the last millennium.} The solid colored lines represent the ensemble mean, black solid lines denote the 30-year running means, dark shading the interquartile range, and light shading the central 99\% confidence interval. The red dashed line in (\textbf{a}) is the 30-year running mean of sea-ice area from \citet{brennan2022reconstructing}. Light orange shading denotes the MCA and light blue shading the LIA.}
    \label{fig:MCA_LIA_ts}
\end{figure*}

Differences between the MCA and LIA are also evident in sea-ice area, sea-ice volume and OHC300 (Fig.~\ref{fig:MCA_LIA_ts}). Specifically, sea-ice area increases around 5\% from 1.1$\times 10^{13}$m$^2$ to 1.15$\times 10^{13}$m$^2$ from the MCA to the LIA. Sea-ice volume increases by about 11\% from 3.5$\times 10^{13}$m$^3$ to 3.9$\times 10^{13}$m$^3$, which we attribute to the longer persistence time of sea-ice volume relative to area \citep{guemas2016review}. Compared to the sea-ice area reconstruction of \citet{brennan2022reconstructing}, we find largely similar centennial-scale results (Fig. \ref{fig:MCA_LIA_ts}a). In particular, both reconstructions show a decline in sea-ice area that began in the early 19th century, and continues to the present. LMR Seasonal has less amplitude on decadal time scales, which is especially evident during the Early Twentieth Century Warming (~1920--1950). We attribute this difference to weak co-variability between 2-m air temperature and sea-ice in the CCSM4 LM simulation, which \citet{brennan2022reconstructing} rectified with covariance inflation; here we do not use covariance inflation. Finally, we note that differences in OHC300 show a decrease from the MCA to LIA of about 1--1.5$\times 10^{8}$ J/m$^2$, or an average of about 10 mW/m$^2$. We note that the overall trend in ocean heat content is similar to that shown in \citet{gebbie2019little}.

\section{Discussion and Conclusions} \label{sec:conclusion}

This study introduces LMR Seasonal, a new reconstruction of coupled atmosphere--ocean--sea-ice climate variability over the Last Millennium, using a novel seasonal ``online" data assimilation method and a new seasonal update strategy. The reconstruction is skillful in both space and time when compared with instrumental observations across the climate variables considered. Skill is primarily attributed to the efficient utilization of proxy information, allowing for updates to model forecasts during assimilation that accurately reflect seasonal variability in the proxies. Additionally, verification against independent (non-assimilated) proxies shows the robustness of the reconstruction in the pre-instrumental period. 

We used the new reconstruction to examine two key measures of climate variability over the last millennium: ENSO and pre-instrumental trends related to the transition from the Medieval Climate Anomaly to the Little Ice Age. For ENSO, LMR Seasonal is able to accurately capture the space--time evolution of tropical SST for four different ENSO categories during the 20th century. Given the large increase in sample size of ENSO over the last millennium compared to the 20th and 21st centuries, LMR Seasonal potentially offers a new resource for ENSO research. For temperature trends of the last millennium, we find that LMR Seasonal captures seasonal variability consistent with orbital forcing, including polar amplification. Moreover, LMR Seasonal demonstrates a pronounced distinction between the Medieval Climate Anomaly (MCA) and the Little Ice Age (LIA), consistent with established climatological studies. This distinction is significantly enhanced by the seasonal updating scheme, which ensures that summer-biased proxies do not dilute the cooler signatures of other seasons during the LIA.

While the LMR Seasonal framework is not specifically optimized for sea-ice reconstruction, it nonetheless demonstrates reasonable skill, compared to previous studies in capturing large-scale sea-ice variability , justifying the inclusion of sea ice in our coupled DA system.

Future studies could expand upon this work by incorporating additional proxy data and exploring regional climate events during the Last Millennium with much larger samples than are available with instrumental reanalyses. Moreover, extending this approach to reconstructions at finer spatial resolution could provide deeper insights into regional climate phenomena and their global implications.

\acknowledgments
This research was supported by NSF awards 2402475, 2202526 and 2105805, and Heising-Simons Foundation award 2023-4715. We thank the editor and all three reviewers for their constructive and insightful comments, which have greatly improved the manuscript. We appreciate comments and conversations related to this work with Feng Zhu (NSF National Center for Atmospheric Research), Olivia Truax (University of Canterbury), Julien Emile-Geay (University of Southern California), Dominik Stiller (University of Washington) and Vince Cooper (University of Washington). 

\datastatement
The PSM package ``CFR" \citep{zhu2023pseudoproxy,zhu2024cfr} and PAGES2K database are located at  \url{https://github.com/fzhu2e/cfr}. The plotting package SACPY \citep{meng2021research,meng2023sacpy} is located at \url{https://github.com/ZiluM/sacpy}. The LMR Seasonal reconstruction code are located at \url{https://github.com/ZiluM/LMR_Seasonal} \citep{meng2025coupled}. The LMR Seasonal Reanalysis Dataset is available at \url{https://doi.org/10.5281/zenodo.17268597} for the ensemble mean, and the full dataset can be accessed at \url{https://atmos.uw.edu/\string~zilumeng/LMR_Seasonal/}.    

Supplementary Information is available at \url{https://zilum.github.io/paper/LMR_Season-2.pdf} and is also supplied as an arXiv ancillary file.

\FloatBarrier
\balance
\bibliographystyle{ametsocV6}
\bibliography{references}

\end{document}